\documentclass[preprint,3p,times,authoryear]{elsarticle}%[preprint,review]

\usepackage[utf8]{inputenc}
\usepackage{amssymb}
\usepackage{amsmath} 
\usepackage{epstopdf}       
\usepackage{flushend}      
\usepackage[colorlinks = true, linkcolor = blue, urlcolor  = blue, citecolor = blue, anchorcolor = blue]{hyperref}
\usepackage{color,soul}

\journal{ }

\usepackage[super]{nth}

\newcommand{\E}{\mathrm{E}}

\newcommand{\Var}{\mathrm{Var}}

\newcommand{\ie}{\textit{i.e.}}
\newcommand{\eg}{\textit{e.g.}}

\newcommand{\taup}{\tau_\mathrm{p}}
\newcommand{\hattaup}{\hat\tau_\mathrm{p}}
\newcommand{\tildetaup}{\tilde\tau_\mathrm{p}}
\newcommand{\taupmin}{\tau_\mathrm{p}^\mathrm{min}}
\newcommand{\taupmax}{\tau_\mathrm{p}^\mathrm{max}}
\newcommand{\taui}{\tau_0}
\newcommand{\dtaui}{\partial_t \tau_0}
\newcommand{\taur}{\tau_\mathrm{r}}
\newcommand{\taucr}{\tau_\mathrm{cr}}
\newcommand{\taukin}{\tau_\mathrm{kin}}
\newcommand{\taucrpred}{\tau_\mathrm{cr}^\mathrm{pred}}
\newcommand{\taucrsim}{\tau_\mathrm{cr}^\mathrm{sim}}
\newcommand{\taun}{\tau_\mathrm{n}}
\newcommand{\xnuc}{x_\mathrm{cr}}
\newcommand{\xnucpred}{x_\mathrm{cr}^\mathrm{pred}}
\newcommand{\xnucsim}{x_\mathrm{cr}^\mathrm{sim}}

\newcommand{\dc}{d_\mathrm{c}}
\newcommand{\slip}{\delta}
\newcommand{\sliprate}{\partial_t \delta}
\newcommand{\fracE}{\Gamma}
\newcommand{\weakrate}{W}

\newcommand{\sysL}{L}
\newcommand{\taustr}{\tau_\mathrm{f}}
\newcommand{\hn}{h_\mathrm{n}}

\newcommand{\argmin}{\mathrm{arg} \min}
\newcommand{\shearmodulus}{G}
\newcommand{\equivshearmodulus}{G^*}
\newcommand{\poisson}{\nu}
\newcommand{\density}{\rho}
\newcommand{\freq}{k}

\newcommand{\corlen}{\xi_0}

\newcommand{\staticERR}{G_\mathrm{II}}
\newcommand{\Dtau}{\Delta \tau}

\newcommand{\cracklength}{h}

\begin{document}

\begin{frontmatter}






\title{Stochastic Properties of Static Friction}%: The Role of Spatial Correlation}


\author[First,Second]{Gabriele Albertini}

\author[First]{Simon Karrer}
%\ead{karrers@student.ethz.ch}

\author[Second]{Mircea D. Grigoriu}
%\ead{mdg12@cornell.edu}

\author[First]{David S. Kammer}
\ead{dkammer@ethz.ch}

\address[First]{Institute for Building Materials, ETH Zurich, Switzerland}
\address[Second]{Civil and Environmental Engineering, Cornell University, Ithaca, NY, USA}

\begin{abstract}

The onset of frictional motion is mediated by rupture-like slip fronts, which nucleate locally and propagate eventually along the entire interface causing global sliding. 
The static friction coefficient is a macroscopic measure of the applied force at this particular instant when the frictional interface loses stability. 
However, experimental studies are known to present important scatter in the measurement of static friction; the origin of which remains unexplained.
Here, we study the nucleation of local slip at interfaces with slip-weakening friction of random strength and analyze the resulting variability in the measured global strength. 
Using numerical simulations that solve the elastodynamic equations, we observe that multiple slip patches nucleate simultaneously, many of which are stable and grow only slowly, but one reaches a critical length and starts propagating dynamically. 
We show that a theoretical criterion based on a static equilibrium solution predicts quantitatively well the onset of frictional sliding.
We develop a Monte-Carlo model by adapting the theoretical criterion and pre-computing modal convolution terms, which enables us to run efficiently a large number of samples and to study variability in global strength distribution caused by the stochastic properties of local frictional strength. 
The results demonstrate that an increasing spatial correlation length on the interface, representing geometric imperfections and roughness, causes lower global static friction. 
Conversely, smaller correlation length increases the macroscopic strength while its variability decreases.
We further show that randomness in local friction properties is insufficient for the existence of systematic precursory slip events. 
Random or systematic non-uniformity in the driving force, such as potential energy or stress drop, is required for arrested slip fronts.
Our model and observations provide a necessary framework for efficient stochastic analysis of macroscopic frictional strength and establish a fundamental basis for probabilistic design criteria for static friction.
%The proposed model provides a necessary framework for efficient stochastic analysis of macroscopic frictional strength and establishes a fundamental basis for probabilistic design criteria for static friction.
\end{abstract}


\end{frontmatter}

%\linenumbers

\section{Introduction}\label{sec:introdution}

Static friction is the maximal shear load that can be applied to an interface between two solids before they start to slide over each other. The famous Coulomb friction law \citep{amontons_resistance_1699,coulomb_theorie_1785,popova_research_2015} states that static friction is proportional to the normal load with the friction coefficient being the proportionality factor. The friction coefficient is generally reported as function of the contacting material pair, which is often misinterpreted as the friction coefficient being a material (pair) property. While proportionality of friction to normal load is mostly valid, the friction coefficient is geometry-dependent and thus varies for different experimental setups with the same material pair~\citep{ben-david_static_2011}. The underlying cause for this observation is the mechanism governing the onset of frictional sliding, which has been shown to be a fracture-like phenomenon~\citep{svetlizky_classical_2014,svetlizky_dynamic_2020,rubino_understanding_2017}. The geometry and deformability of the solids lead to a non-uniform stress state along the interface. As a consequence, local frictional strength is reached at a critical point and slip nucleation starts, from where it extends in the space-time domain -- just like a crack -- until the entire interface transitioned and global sliding occurs. 

Variations in the static friction force, however, do not only occur because of changes in the loading configuration. Experiments have shown that the measured friction force varies also from one experiment to another when the exact same setup and exact same specimens are used. For instance, \citet{rabinowicz_friction_1992} showed that the static friction coefficient of a gold-gold interface, measured by a tilting plane friction apparatus, varies from $0.32$ to $0.80$ for normal load $75~\mathrm{g}$. Similar but to a lesser extent, \citet{ben-david_static_2011} also observed variations in the static friction coefficient of glassy polymers when the loading configuration was fixed.
%With varying normal load the mean coefficient of friction is 0.98 for 0.01g and 0.45 for 300g.

While these variations are not often reported, they are an important factor in the absence of a complete and consistent theory for friction~\citep{spencer_cutting_2015}. If (seemingly) equivalent experiments lead to a large range of observations without consistent trends, it is challenging to isolate the relevant from the irrelevant contributions and, therefore, nearly impossible to create a fundamental understanding of the underlying process. Even though the presence of these large variations has important implications for the study of friction, current knowledge about the origin and properties of these observed variations in macroscopic friction remains limited.

One possible origin is randomness in local friction properties. Interfaces have been shown to consist of an ensemble of discrete micro-contacts \citep{bowden_friction_1950,dieterich_direct_1994,sahli_evolution_2018}, which are created by surface roughness \citep{thomas_rough_1999,hinkle_emergence_2020} when two solids are brought into contact. This naturally leads to a system with random character, where micro-contacts of random size are distributed randomly along the interface \citep{greenwood_contact_1966,persson_theory_2001,hyun_elastic_2007,yastrebov_infinitesimal_2015}. Since frictional strength is directly related to the cumulative contact area of theses micro-contacts \citep{bowden_friction_1950,greenwood_contact_1966}, and the micro-contacts are the result of random surface roughness, the local frictional strength is likely also random. 

Surprisingly, the effect of interfacial randomness on friction remains largely unexplored. Most of previous work is focused on how (random) surface roughness is related to various friction phenomenology including rate-dependence \citep{li_friction_2013,lyashenko_comment_2013}, local pressure excursions within lubricated contact \citep{savio_boundary_2016}, chemical aging \citep{li_chemical_2018}, or the existence of static friction \citep{sokoloff_static_2001}. How interfacial randomness causes variation in these observations has, however, not been studied so far. Only most recently, \citet{amon_avalanche_2017} and \citet{geus_how_2019} have considered variability in friction. 
\citet{amon_avalanche_2017} showed that systems with a nonuniform initial stress state with long range coupling are characterized by two regimes: at low loading, small patches of the system undergo sliding in an uncorrelated fashion; at higher loading, instabilities occur at regular intervals over patches of increasing size -- just like confined stick-slip events \citep{kammer_linear_2015,bayart_fracture_2016} -- and eventually span the whole system.
\citet{geus_how_2019} simulated interface asperities as an elasto-plastic continuum with randomness in its potential energy and show that the stress drop during a stick-slip cycle is a stochastic property which vanishes with increasing number of asperities. These results demonstrate well the stochastic character of macroscopic friction due to random interface properties. However, the effects of interfacial randomness on the variability of macroscopic static friction, \eg, the friction coefficient, has not been studied yet. 

% discorder at asperity level and long-range elastic interactions, and inertia
% they study effect of system size and find A_c ~ N^0.32 with \sigma_n - \sigma_c ~ N^-0.16
% the critical stress is stochastic and stick slip vanishes with increasing N
%\citep{geus_how_2019}

% precursors to avalanches in frictional model to study the regularity of stick-slip mechanism.
%\citep{amon_avalanche_2017}

% one-dimensional elastomer on randomly rough rigid surface (self affined fractal).
% kinetic friction force increase first, plateaus and then decreases again with rupture velocity
% lyshenko shows that li system is completely unrealistic
%\citep{li_friction_2013,lyashenko_comment_2013}

% showed existence of static friction for distribution of asperities at multimicron length scales, because they are always in strong pinning regime.
%\citep{sokoloff_static_2001}

% regular and random slip patterns on a surface lead to pressure excursions within a lubricated contact
%\citep{savio_boundary_2016}

% chemical aging for randomly rough contacts has a logarithmic dependence on time. Surface roughness affects the aging behavior primarily by modifying the real contact area and the local contact pressure
%\citep{li_chemical_2018}

% experiments
% the real contact area decreases under shear, [...], starting well before macroscopic sliding.
%\citep{sahli_evolution_2018}

Here, we address this gap of knowledge and aim at a better understanding of the stochastic properties of static friction. We present a combined numerical and theoretical study that links randomness of local friction properties with observed variability in macroscopic strength. Using dynamic simulations, we will show that the macroscopic friction threshold is attained when a local slipping area, of which many can co-exist, reaches a critical length and nucleates the onset of friction. This nucleation patch becomes unstable and propagates across the entire interface causing global sliding. We will then show that a quasi-static equilibrium theory, which takes an integral form, predicts quantitatively well the critical stress level that causes nucleation of global sliding. Based on this theoretical model, we will develop fast and accurate Monte Carlo simulations using a Fourier representation of the integral equations, and demonstrate the extent of variability in macroscopic static friction based on random interface strength with various correlation lengths. Finally, we will show that a decreasing interfacial correlation length leads to higher macroscopic strength with decreased variability.

This paper is structured as follows. First, we provide a problem statement in Sec.~\ref{sec:probstatement} including a description of the physical system, the stochastic properties, and our approach to generate random strength fields. In Sec.~\ref{sec:dynamicsim}, we present the numerical method used to simulate the onset of frictional sliding and compare simulation results of critical stress leading to global sliding with predictions based on a theoretical model. This model is then used in an analytical Monte Carlo study, which is developed and presented in Sec.~\ref{sec:analyicalMCstudy}. The implications of our model assumptions as well as the model results are discussed in Sec.~\ref{sec:discussion}. Finally, we provide a conclusion in Sec.~\ref{sec:conclusion}.

\section{Problem Statement}\label{sec:probstatement}

In this section, we first provide a description of the physical problem that we consider throughout this paper. We then describe the stochastic properties of the strength profile along the interface and, finally, explain how we generate these random fields. 

\begin{figure}
\begin{center}
  \includegraphics[width=3.375in]{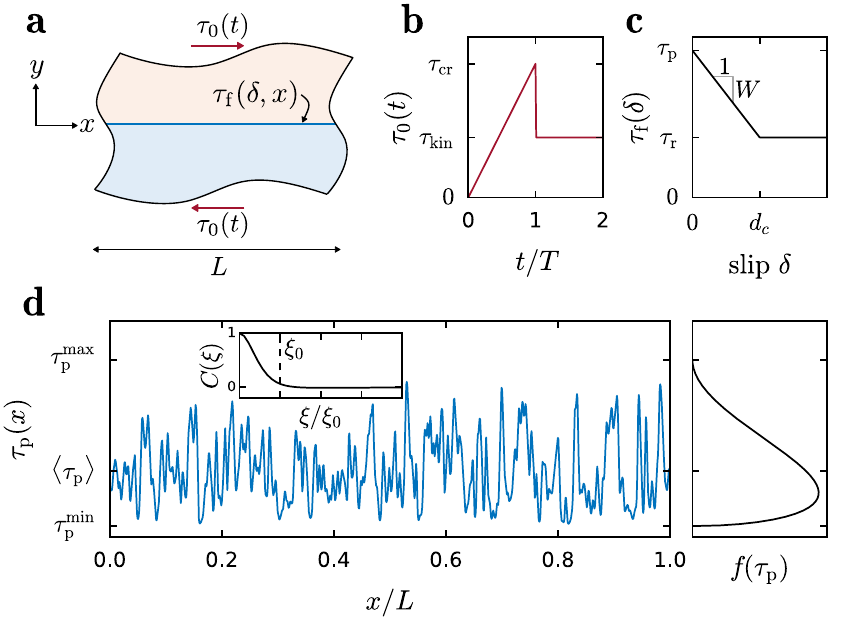}
  \caption{Problem statement. 
  (a)~A frictional interface (blue line) of strength $\tau_\mathrm{f}(\delta,x)$ embedded within two semi-infinite elastic solids, which are periodic in $x$ with period $\sysL$ and infinite in $y$. A uniform loading $\taui(t)$ is applied. 
  (b)~$\taui(t)$ is increased linearly with time $t$ up to the onset of frictional motion when the stress drops from its critical value $\taucr$ to a kinetic level $\taukin$.
  (c)~The constitutive relation of the frictional interface is a linear slip-weakening law $\taustr(\delta)$ with random peak strength $\taup(x)$ and constant weakening rate $W$ (see Eq.~\ref{eq:linslipweaklaw}).
  (d)~$\taup(x)$ is a random field with spatial correlation $C(\xi)$ (inset) and probability density function $f(\taup)$ (right).}
  \label{fig:model}  
\end{center}
\end{figure}

\subsection{Physical Problem}\label{sec:physproblem}

We study the macroscopic strength of a frictional interface. Our objective is to provide a fundamental understanding of the effect of local variations in frictional strength on the macroscopic response. For this reason, we focus on the simplest possible problem -- without oversimplifying the constitutive relations of the bulk and the interface.
Specifically, we consider a two-dimensional (2D) system consisting of two semi-infinite elastic solids, as shown in Fig.~\ref{fig:model}a.
The domain is infinite in the $y$ direction and periodic in $x$ with period $\sysL$. 
Both materials have the same elastic properties.

We apply a uniform shear load $\taui(t)$ that increases quasi-statically with time (see Fig.~\ref{fig:model}b). Once $\taui(t)$ reaches the macroscopic strength of the interface $\taucr$, the interface starts to slide and the frictional strength suddenly reduces to its kinetic level $\taukin$. This observed reduction in shear stress is typically associated with friction-weakening processes, which may depend on various properties, such as slip, slip rate, and interface state. The critical shear stress $\taucr$, if divided by the contact pressure, corresponds to the static macroscopic friction coefficient. Similarly, $\taukin$ is proportional to the kinetic friction coefficient.

These macroscopic observations depend on the local interface properties, which are the peak strength $\taup(x)$, and residual strength $\taur(x)$. As we will show, the local properties are generally different from the macroscopic properties; particularly, in the case of non-uniform stress or strength. 
%the macroscopic strength is generally different from the local peak strength $\taup(x) \neq \taucr$; particularly, in the general case of non-uniform stress or strength. However, the macroscopic residual strength is a good indicator of the local value $\taur$ since the stress states becomes often becomes uniform. The local evolution of the frictional strength 
For simplicity, we describe the evolution of the local frictional strength as a linear slip-weakening law, which is shown in Fig.~\ref{fig:model}c and is given by 
\begin{equation}
    \taustr (\slip) = \taur + \weakrate (\dc - \slip) H(\dc - \slip) ~,
    \label{eq:linslipweaklaw}
\end{equation}
where $\slip(x)$ is local slip, $\dc(x)$ is a characteristic length scale, and $\weakrate(x) = (\taup(x) - \taur(x)) / \dc(x)$ is the weakening rate. $H(.)$ is the Heaviside function.
%$\tau_s(\delta)=\max(\tau_p - W \delta,\tau_r)$, where $W=\taup/\delta_c$

We consider a heterogeneous system with local peak strength $\taup(x)$ being a random field, as further described in Sec.~\ref{sec:stochprop}. To reduce complexity of the problem, we assume uniform residual strength\footnote{We use $\partial_i$ as short notation for partial derivative with respect to $i$.} $\partial_x \taur = 0$ and uniform weakening rate $\partial_x W = 0$. The variation in local peak strength is thought to represent possible heterogeneity in the material, but also the effect of surface roughness, which leads to a real contact area that consists of an ensemble of discrete contact points with varying properties. The implications of this approach will be discussed in depth in Sec.~\ref{sec:discussion}. 

\subsection{Stochastic Properties of Frictional Interface}\label{sec:stochprop}
\begin{figure}
\begin{center}
  \includegraphics[width=4.5in]{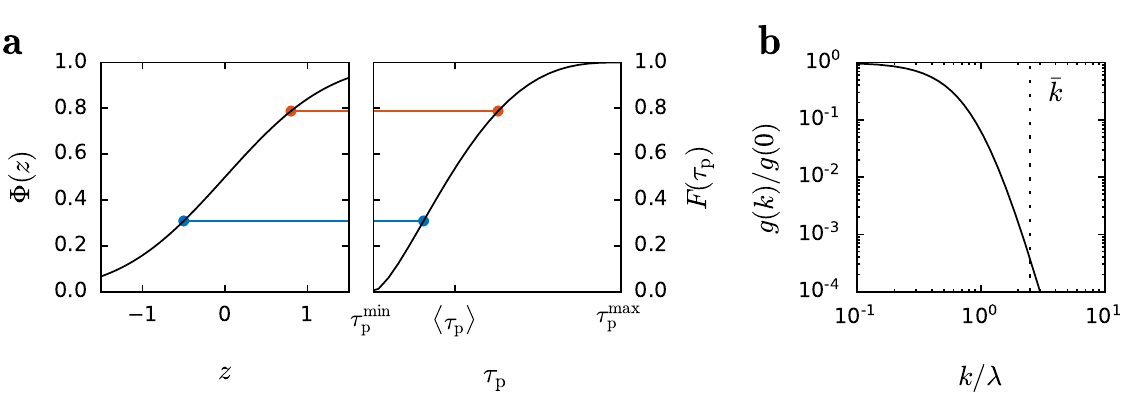}
  \caption{Stochastic properties of the random peak strength field $\taup(x)$. 
    (a)~The random variable $\taup$ is generated by applying a nonlinear mapping $F^{-1}\circ \Phi(x)$ (Eq.~\ref{eq:stoch_process}) onto the Gaussian random variable $z$. Two colored dots inked by a line represent a $(z,\taup)$ pair with equal cumulative density $F(\taup)=\Phi(z)$, which is the criterion imposed by the nonlinear mapping.
    (b)~Normalized spectral density function $g(\freq)/g(0)$ (Eq.~\ref{eq:markov4}).
    $\bar \freq$ is the truncation frequency and $\lambda=2\pi/\corlen$ is the cutoff frequency.}
  \label{fig:stoch_properties}  
\end{center}
\end{figure}

The local peak strength $\taup(x)$ is modeled as a stationary non-Gaussian random field with specified cumulative distribution function $F(\taup)$ and corresponding probability density $f(\taup)$, as shown in Fig.~\ref{fig:model}d. 
The random field is defined by the nonlinear mapping
\begin{equation}
\label{eq:stoch_process}
\taup(x) = F^{-1}\Big({\Phi\big(z(x)\big)}\Big) ~,
\end{equation}
where $z(x)$ is Gaussian with zero mean and unit variance and $\Phi$ its cumulative distribution, depicted in Fig.~\ref{fig:stoch_properties}a-left.
$F$ and $\Phi$ are monotonic by definition, so their inverse exist, which can be used to prove that $P\left(\taup(x)\leq \tau\right) = F(\tau)$.
Further, with $\taup$ being the local peak strength of the interface, it needs to satisfy some physical requirements. First, the peak strength is always higher than the residual strength, \ie, $\taup^{\min}\geq \taur$. Second, it maximum value is limited by the material properties. For this reason, we require that $\taup \in (\taup^{\min}, \taup^{\max} )$, which we achieve by setting $F(\taup)$ as a Beta cumulative distribution function (see Fig.~\ref{fig:stoch_properties}a-right). 

The spatial evolution of $z(x)$ is specified by its power spectral density $g(\freq)$, which corresponds to the Fourier transform of the correlation function $C_z(\xi)$, \ie,
\begin{equation}
g(\freq)  \equiv \int_{-\infty}^{+\infty} C_z(\xi)e^{-i \freq\xi}\mathrm{d}\xi ~,
\end{equation}
where $\freq$ is the angular wave number.
We assume that $z(x)$ has a power spectral density
\begin{equation}
\label{eq:markov4}
    g(\freq) \propto {(\freq^2+\lambda^2)^{-4}} ~,
\end{equation}
where $\lambda$ is the cutoff frequency, above which the spectral density decays as a power law $\sim k^{-8}$ (see Fig.~\ref{fig:stoch_properties}b). 
The correlation length $\corlen$ is a measure of memory of the random field; the longer $\corlen$ the longer the memory.
$\corlen$ is inversely proportional to $\lambda$, and we define\footnote{The correlation length does not have a precise definition. And alternative definition is  $C(\corlen) = \exp(-1)$.} it as $\corlen=2\pi/\lambda$. 
Since the correlation function of $z$ is positive, $C_z(\xi)>0$, it is not greatly affected by the nonlinear mapping $F^{-1}\circ \Phi$ and $C_z(\xi) \approx C_{\taup}(\xi)$ \citep[p.48]{grigoriu_applied_1995} and features such as $\corlen$ are preserved (see Fig.~\ref{fig:model}.d inset). 
The assumption of using this specific spectral density and probability distribution are discussed in Sec.~\ref{sec:impl_mc}.
%Note that the corresponding  Hurst exponent for the given power spectral density is $H=3$ \citep{mai_spatial_2002}. does not apply because 0<H<1

% if fractal  $g(\freq)\propto (\freq^2)^{-4+D}$ where $D=E+1-H$ is the fractal dimension, $E$ the Euclidean dimension.

\subsection{Random Field Samples Generation}\label{sec:randgeneration}

The samples of random field $\taup(x)$ are generated as follows. First, the Gaussian random field $z(x)$ is generated using a spectral representation
\begin{equation}
\label{eq:gaussian_process}
    z(x) =\sum_{j=1}^J \sigma_j \left(A_j \cos(\freq_j x) + B_j \sin(\freq_j x)\right),
\end{equation}
where $A_j$ and $B_j$ are independent Gaussian random variables with zero mean and unit variance and modal angular wave-number is $\freq_j=2\pi j/L$. 
The fundamental wavelength of the field  $2\pi/\freq_1=L$ is chosen such that it corresponds to the domain size $L$, which implies that $z(x)$ is periodic over $L$, and so is $\tau_p(x)$.
The modal variance $\sigma_j^2\propto g(\freq_j)$ corresponds to the discrete spectral density, which is normalized to assure that $z$ has unit variance %($\sum \sigma_j^2=1$)
\begin{equation}
    \sigma_j^2={\frac{g(\freq_j)}{\sum_{j=1}^{J} g(\freq_j)}}.
\end{equation}
Due to the discrete representation of $z(x)$, we apply a truncation frequency that is considerably larger than the cutoff frequency $\bar \freq\equiv\freq_J = 2.5 \lambda$. This ensures that most of the spectral power is preserved:
\begin{equation}
\frac{\int_0^{\bar\freq}g(\freq) \mathrm{d}\freq}{\int_0^\infty g(\freq)\mathrm{d}\freq}\approx0.9997
\end{equation}
Further increase in $\bar \freq$ would include additional high frequency modes but with negligible amplitudes.
Finally, once $z(x)$ has been generated, we apply the nonlinear mapping $F^{-1}\circ \Phi$ (Eq.~\ref{eq:stoch_process} and visualized in Fig.~\ref{fig:stoch_properties}a) and obtain the random field $\taup(x)$.
Fig.~\ref{fig:model}d shows a sample of $\taup(x)$ generated using the described procedure with corresponding correlation function and probability density. 

\section{Dynamic Simulations}
\label{sec:dynamicsim}

In the following, we will first present the numerical method and model setup applied in our simulations of the onset of friction. We then provide a theoretical model to describe the simulations and present a comparison between the theoretical predictions with the numerical results.

\subsection{Numerical Method}\label{sec:nummethod}

We model the physical problem, as described in Sec.~\ref{sec:physproblem}, with the Spectral Boundary Integral Method (SBIM) \citep{geubelle_spectral_1995,breitenfeld_numerical_1998}. This method solves efficiently and precisely the elasto-dynamic equations of each half space.
The spectral formulation applied in SBIM naturally provides periodicity along the interface. The half spaces are perfectly elastic and we apply a shear modulus of $\shearmodulus = 1~\mathrm{GPa}$, Poisson's ratio of $\poisson = 0.33$ and density $\density = 1170~\mathrm{kg/m}^3$, and impose a plane-stress assumption. While we will report our results in adimensional quantities, we note that these parameters correspond to the static properties of glassy polymers, which have been widely used for friction experiments \citep{svetlizky_classical_2014,rubino_understanding_2017}.
%We model the physical problem, as described in Sec.~\ref{sec:methods}, by solving the elasto-dynamic equations for solids. We apply the Spectral Boundary Integral Method (SBIM) [CITE], which solves efficiently and precisely the elasto-dynamic equations of half spaces. Two half spaces are coupled by an interface friction law. 
\begin{figure}
\begin{center}
  \includegraphics[width=5in]{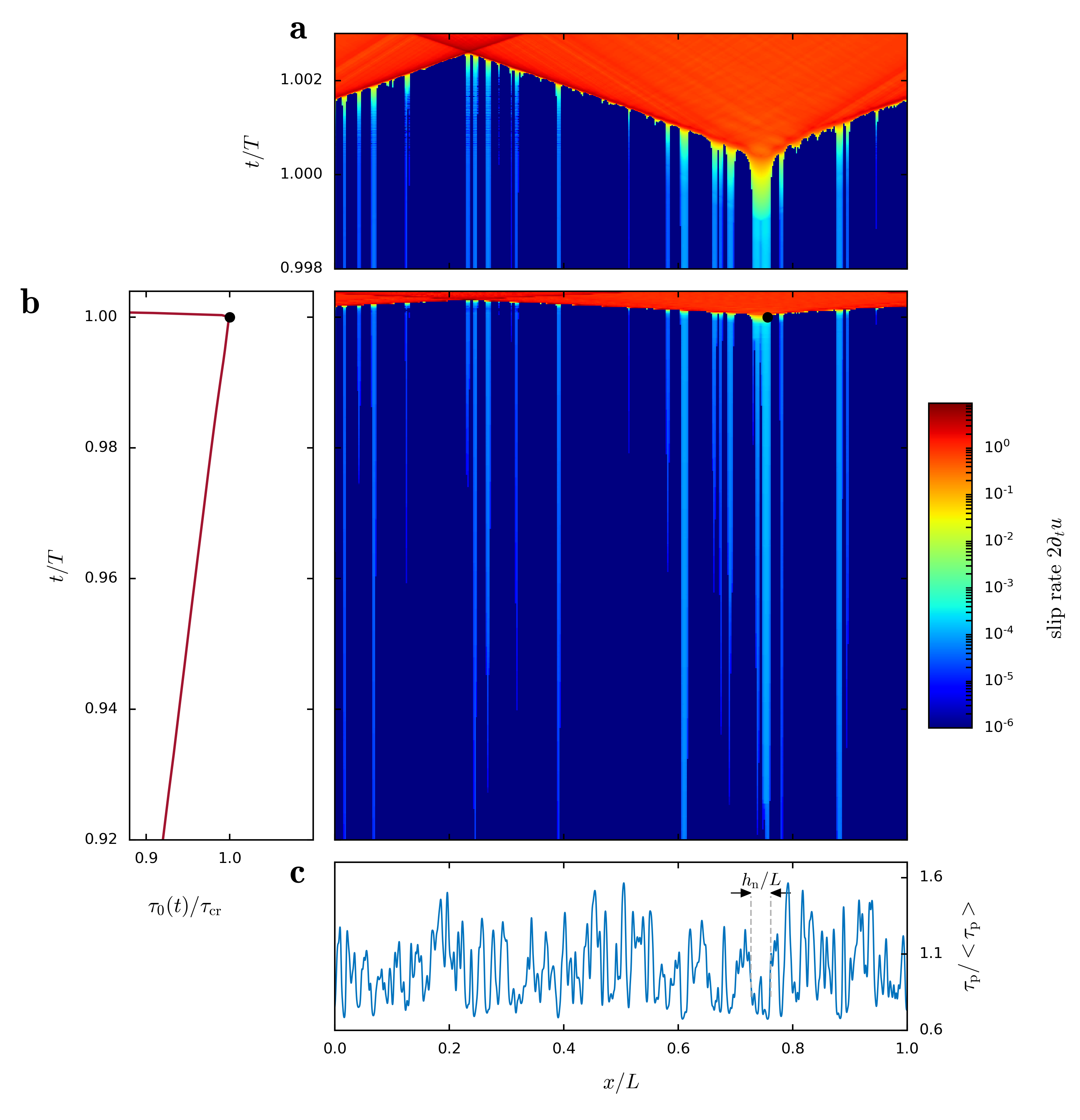}
  \caption{Representative numerical simulation of the onset of frictional sliding along an interface with random peak strength. (a)~Space-time diagram of slip rate along the interface. Time is normalized by the time of friction onset $T$. Nucleation occurs at $x/L \approx 0.75$. (b-right)~Same space-time diagram as in (a) with larger time span. Nucleation is marked by a black dot at $x/L \approx 0.75$ and $t/T = 1$. (b-left)~Evolution of applied load $\taui(t)$ normalized by its maximum value $\taucr$. This corresponds to Fig.~\ref{fig:model}b. (c)~Random profile of peak strength $\taup(x)$ for simulation shown in (a-b). The correlation length is $\corlen/\hn = 0.25$. The size of the critical nucleation patch $\hn$ is marked by black arrows.}
  \label{fig:simulation}  
\end{center}
\end{figure}

The interface between the two half spaces is coupled by a friction law as given by Eq.~\ref{eq:linslipweaklaw}. The friction law corresponds essentially to a cohesive law, as known from fracture-mechanics simulations, but applied to the tangential direction. It describes the evolution of local strength as a function of slip. We apply peak strength $\taup(x)$ as a random field, following the description provided in Sec.~\ref{sec:stochprop}, and constant $\taur$. 
$\taup(x)$ follows a Beta distribution with $\alpha=1.5$ and $\beta=3$. 
We impose a maximum value for relative peak strength of $\max(\taup(x)-\taur)=1.66 ~\mathrm{MPa}$ and minimum value of $\min(\taup(x)-\taur)=0.66~\mathrm{MPa}$. Therefore, the random field has a mean value of $\left <\taup-\taur\right>=1~\mathrm{MPa}$ and standard deviation of $0.2~\mathrm{MPa}$. We further apply a constant slip-weakening rate of $W=0.5~\mathrm{TPa/m}$, which is representative for glassy polymers \citep{svetlizky_dynamic_2020}. Finally, a slowly increasing uniform stress $\taui(t)$ is applied along the entire interface. 
% we do not need the following
%\begin{itemize}
%    \item standard deviation $\sqrt{\left<\left(\taup-\left<\taup\right>\right)^2\right>}=0.2$ MPa % don't need formula
%    \item $\taur=0$ MPa %don't need value
%    \item $\left< \Gamma \right>=1$ J/m$^2$ %don't need fracture energy
%\end{itemize}

We use a repetition length of $\sysL = 0.1~\mathrm{m}$, which is, as we will show, considerably larger than the characteristic nucleation length scale. The interface is discretized by $512 - 1024$ nodes. We verified convergence with respect to discretization, loading rate, and time step. 

The results of a representative simulation are shown in Fig.~\ref{fig:simulation}. 
%The applied random profile for $\taup(x)$ is depicted in Fig.~\ref{fig:simulation}c. 
%The evolution of local slip rate is shown in the space-time color plot in Fig.~\ref{fig:simulation}b-right. 
The $\taup(x)$ profile has many local minima (see Fig.~\ref{fig:simulation}c). Depending on their value, these minima cause localized slip, as evidenced by bright blue vertical stripes over most of the time period shown in Fig.~\ref{fig:simulation}b-right. These localized slip patches grow slowly with increasing loading, which is difficult to see for most patches in Fig.~\ref{fig:simulation}b-right. Growth is easiest observed for the slip patch at $x/\sysL \approx 0.75$. Incidentally, this patch grows enough to reach a critical size from which on the patch becomes unstable, marked by a black dot, and starts growing dynamically. This dynamic propagation, see orange-red area in Fig.~\ref{fig:simulation}a enlarged from Fig.~\ref{fig:simulation}b-right, does not stop and, therefore, causes sliding along the entire interface -- hence global sliding. The effect on the macroscopic applied force is shown in Fig.~\ref{fig:simulation}b-left, where $\taui(t) = \int_{\sysL}\tau(x,t) \mathrm{d} x$. At the precise moment when the slip patch becomes unstable, marked by a black dot, $\taui(t)$ starts decreasing rapidly. The maximum value, denoted $\taucr$, represents the macroscopic strength of the interface. 

The simulation shows that macroscopic strength is not reached when the first point along the interface starts sliding but when the most critical slip patch becomes unstable, starts propagating dynamically, and "breaks" the entire interface. Therefore, the criterion determining macroscopic strength is non-local and depends on the stability of local slip patches. In the following section, we will present a theoretical description of this nucleation process and provide a criterion for the limit of macroscopic strength.

\subsection{Theory for Nucleation of Local Sliding}
\label{sec:nuctheory}

During the nucleation process, a weak point along the interface starts sliding. Due to local stress transfer, the size of this slipping area grows continuously until it reaches a critical size and unstable interface sliding occurs \citep{campillo_initiation_1997}. In this section, we will adapt the criterion developed by \citet{uenishi_universal_2003}, which is shortly summarized in \ref{sec:appendixA}, to describe and predict the limits of stable slip-area growth. \citet{uenishi_universal_2003} considered a similar system with two main differences to the problem studied here. First, in their case, the interface strength is uniform and the applied load is non-uniform. \ref{sec:appendixA} shows that both problems result in the same equation for the problem statement and thus lead to the same nucleation criterion. Second, \citet{uenishi_universal_2003} considered a system with an isolated non-uniformity in the applied load. In other words, the applied stress was mostly uniform but with one well-contained local increase. Therefore, the location of nucleation is known in advanced. In our system, where the non-uniform property is random, the location is unknown. We will address this difference here and discuss it further in Sec.~\ref{sec:discussion}.

\citet{uenishi_universal_2003} showed that on interfaces governed by linear slip-weakening friction (Eq.\ref{eq:linslipweaklaw}), there is a unique critical length for stable growth of the slipping area, which can be approximated by
\begin{equation}
    \hn \approx 1.158 \frac{\equivshearmodulus}{\weakrate} ~,
    \label{eq:critlength}
\end{equation}
where $\equivshearmodulus = \shearmodulus / (1-\poisson)$ for mode II plane-stress ruptures, assuming the stress within $\hn$ has not attained the residual value $\taur$ anywhere.
Eq.~\ref{eq:critlength} shows that $\hn$ depends only on the shear modulus $\equivshearmodulus$ and the slip-weakening rate $\weakrate$. Most importantly, the critical length is independent of the shape of the non-uniformity in the system. 
Specifically to our case, it does not depend on the functional form of $\taup(x)$. Since we have homogeneous elastic solids and a uniform slip-weakening rate $\weakrate$, the critical size $\hn$ is unique and uniform along the entire interface. 

The important question for our problem, however, is to determine the level of critical stress that causes a nucleation patch to reach $\hn$ and initiate global sliding. The solution for the stress level leading to nucleation, as derived by \citet{uenishi_universal_2003}, is given by Eq.~\ref{eq:apxAcrittau}, and can be rewritten in terms of $\taup(x)$ and $\hn$ as 
\begin{equation}
    \taun(x) \approx 0.751 \int_{-1}^{+1} \taup\left( \frac{\hn}{2} s+ x\right) ~ v_0(s) ~ \mathrm{d} s ~,
    \label{eq:taun}
\end{equation}
where $x$ is the center location of the nucleation patch and $v_0(s) \approx (0.925-0.308s^2) \sqrt{1-s^2}$ is the first eigenfunction of the elastic problem. 
Note that the transformation applied to the argument of $\taup(x)$ results in the integral being computed over the critical nucleation patch size $\hn$. Eq.~\ref{eq:taun} shows that the nucleation stress, which leads to a nucleation patch of size $\hn$, does clearly depend on the shape of $\taup(x)$.  
Note that Eq.~\ref{eq:taun} assumes that stresses within the nucleation patch have not attained the residual strength yet \emph{i.e.}, the fracture process zone spans the entire crack. In this case, the assumption of small-scale yielding does not hold, and, thus, the Griffith criterion for crack propagation does not apply.

As stated earlier, the nucleation stress $\taun(x)$ was derived for a contained non-uniformity, for which we know the location. Therefore, $\taun(x)$ corresponds to the critical stress of the system. In our system, however, $\taup(x)$ is random and multiple nucleation patches might slowly grow. Determining the critical stress $\taucr$ of the system requires computing the nucleation stress $\taun$ for each nucleation patch and identifying the critical one. To address this aspect, we propose to compute Eq.~\ref{eq:taun} as a weighted moving average over the entire interface, and define the critical stress to be its minimum (see Fig.~\ref{fig:theory_vs_sim}a). Therefore, we define the critical stress $\taucr$ as
\begin{equation}
    \taucr = \taun(\xnuc) \quad \mathrm{such ~that} ~ \taucr < \taun(x) \quad \forall x\neq \xnuc ~.
    \label{eq:taucr}
\end{equation}
For simplicity, we refer to this definition also as $\taucr=\min(\taun(x))$ and $\xnuc=\argmin(\taun(x))$. 
While it is possible, but not very likely, to have multiple minima of $\taun$ with the same amplitude, this does not affect the resulting $\taucr$. 
However, multiple $\xnuc$ could coexist which would result in multiple slip patches becoming unstable simultaneously.
By adopting Eq.~\ref{eq:taun} and defining Eq.~\ref{eq:taucr}, we essentially assume that there is no interaction between nucleation patches.
We will verify the validity of this assumption in the following section.

\subsection{Results}\label{sec:results}

We compare the results from numerical simulations, as described in Sec.~\ref{sec:nummethod}, with the theoretical prediction from Sec.~\ref{sec:nuctheory} by analyzing simulations with random $\taup(x)$ generated using the method described in Sec.~\ref{sec:randgeneration}. For each of the three different correlation lengths $\corlen/\hn = 0.25$, $0.5$, and $2.0$ we run $20$ simulations. The system size is fixed and chosen such that it is considerably larger than the nucleation length, \ie, $\hn/\sysL = 0.034$. 

\begin{figure}
\begin{center}
  \includegraphics[width=3.375in]{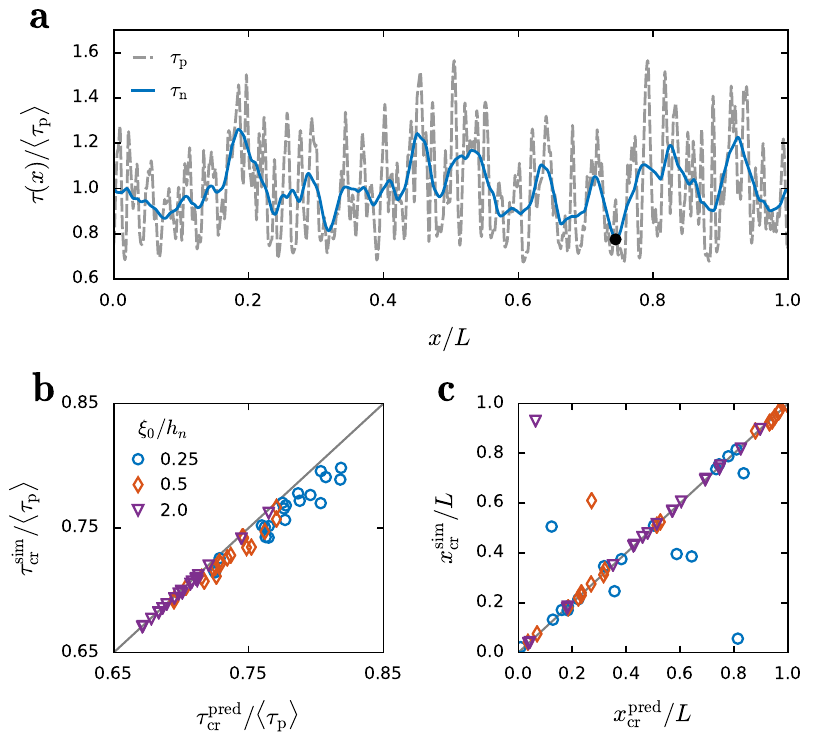}
  \caption{Verification of nucleation criterion on numerical simulations. (a)~The random profile of $\taup(x)$ from simulation shown in Fig.~\ref{fig:simulation} is depicted by dashed gray line. The nucleation stress $\taun(x)$ computed from $\taup(x)$ by Eq.~\ref{eq:taun} is shown as solid blue line. The point of nucleation given by Eq.~\ref{eq:taucr}, \ie, ($\xnuc$,$\taucr$), is marked by a black dot. (b)~Comparison of critical length from theoretical prediction by Eq.~\ref{eq:taucr} $\taucrpred$ as shown in (a) with values measured from numerical simulations $\taucrsim$ as illustrated in Fig.~\ref{fig:simulation}. $20$ simulations are computed for each $\corlen/\hn$ value. (c)~Comparison of nucleation location from theoretical prediction $\xnucpred$ with simulation result $\xnucsim$ for the same $60$ simulations as shown in (b). (b-c)~Gray line indicates slope of $1$.}
  \label{fig:theory_vs_sim}  
\end{center}
\end{figure}

A representative example is shown in Fig.~\ref{fig:simulation}. The size of $\hn/\sysL$ is indicated in Fig.~\ref{fig:simulation}c and appears to provide a reasonable prediction for the nucleation patch size as observed in Fig.~\ref{fig:simulation}a. Further comparison is given in Fig.~\ref{fig:theory_vs_sim}a. First, we illustrate the theoretical prediction. The nucleation stress $\taun(x)$ (solid blue line) is computed from $\taup(x)$ (gray dashed line) using Eq.~\ref{eq:taun}, and $\taucr$ is, according to Eq.~\ref{eq:taucr}, the minimum of $\taun(x)$ (marked by black dot). We find the location of nucleation to be $\xnuc/\sysL \approx 0.75$, which corresponds to our observation from the numerical simulation, as seen in Fig.~\ref{fig:simulation}.

A more precise and systematic comparison is provided in Fig.~\ref{fig:theory_vs_sim}b\&c. We compare the predicted critical stress $\taucrpred$ with the measured value from dynamic simulations $\taucrsim$. We compute $\taucrpred$ as described above with Eq.~\ref{eq:taucr}, and as illustrated in Fig.~\ref{fig:theory_vs_sim}a. We further find $\taucrsim = T \dtaui$, where $\dtaui$ is the applied loading rate and $T$ is the time at which $\taui(t)$ is maximal (see Fig.~\ref{fig:simulation}b-left). Comparison of $\taucrpred$ with $\taucrsim$ is shown in Fig.~\ref{fig:theory_vs_sim}b for all $60$ simulations. The results show that the prediction works generally well. For decreasing $\corlen/\hn$ the prediction becomes slightly less accurate with a tendency to over-predict the critical value. The results further show that the predicted and measured critical stress $\taucr$ increases with decreasing $\corlen/\hn$.

While the location of nucleation is not relevant for the apparent global strength of our system, we compare the predicted and simulated $\xnuc$ for further evaluation of the developed theory. The comparison shown in Fig.~\ref{fig:theory_vs_sim}c uses $\xnucpred$, as given by Eq.~\ref{eq:taucr} and shown for an example in Fig.~\ref{fig:theory_vs_sim}a, and $\xnucsim$ as found by analyzing the simulation data as illustrated in Fig.~\ref{fig:simulation}a\&b-right. The data shows that the prediction works well for most of the simulations. For $8$ simulations, $6$ of which have $\corlen/\hn = 0.25$, the prediction does not work. However, as shown in Fig.~\ref{fig:theory_vs_sim}b, $\taucr$, which is the quantity of interest here, is correctly predicted for all of these cases. The reason for this discrepancies are likely second-order effects, as we will discuss in Sec.~\ref{sec:discussion}.

Overall, the results show that $\taucr$ is quantitatively well predicted by the theory presented in Sec.~\ref{sec:nuctheory}. This allows us to study  systematically the effect of randomness in interface properties by applying the theoretical model in analytical Monte Carlo simulations.

\section{Analytical Monte Carlo Study}
\label{sec:analyicalMCstudy}

In the following section, we introduce Monte Carlo simulations, which are based on the theoretical framework for nucleation of frictional ruptures in a random field of frictional strength $\taup(x)$, as derived in Sec.~\ref{sec:nuctheory}.
The effect of correlation length $\corlen$ on the effective frictional strength $\taucr$  (Eq.~\ref{eq:taucr}), and its probability distribution $f(\taucr)$, is studied, while keeping all other properties constant.
A Monte Carlo study based on the full dynamic problem (Sec.~\ref{sec:nummethod}) would be computationally daunting. 
However, the theoretical framework allows us to evaluate $\taucr$ very efficiently and has been validated by 20 full dynamic simulations for each considered $\corlen$ (see Fig.~\ref{fig:theory_vs_sim}).

\subsection{Monte Carlo Methodology}
The effective frictional strength $\taucr=\min( \taun(x))$ requires the computation of the nucleation strength $\taun(x)$, which involves a convolution of the local peak strength $\taup(x)$ with the eigenfunction $v_0$, given in Eq.~\ref{eq:taun}. 
Considerable computation time can be saved by using a spectral representation of the random field $\taup(x)$:
\begin{equation}
    \label{eq:taup_Fourier}
    \tildetaup(x)=\sum_{j=0}^J \hattaup(\freq_j) e^{-i \freq_j x},
\end{equation}
where $\tilde{}$ signifies that $\tildetaup(x)$ is an approximation of $\taup(x)$ and the number of frequencies $J$ is chosen such that the approximation error $|\tildetaup-\taup|$ is negligible.
$\hattaup(\freq_j)$ is the discrete Fourier transform of $\taup(x)$
\begin{equation}
    \hattaup(\freq_j) = \int_0^L \taup(x) e^{-i \freq_j x}\mathrm{d}x ,
\end{equation}
where $\taup(x)$ is generated using the procedure described in Sec.~\ref{sec:randgeneration}.
By substituting Eq.~\ref{eq:taup_Fourier} into Eq.~\ref{eq:taun} the nucleation strength convolution becomes a dot product:
\begin{equation} 
\label{eq:taun_spec}
\begin{split}
    \tilde\tau_\mathrm{n}(x) &\approx 0.751\int_{-1}^{+1} \sum_{j=0}^J \hattaup(\freq_{j})e^{-i\freq_j\left(s~\hn/2 + x\right)} v_0(s) \mathrm{d} s\\
    &\approx  0.751\sum_{j=0}^J \hattaup(\freq_{j})g_j(x)
\end{split}
\end{equation}
where $g_j(x)=\int_{-1}^{+1}e^{-i\freq_j\left(\frac{\hn}{2} s + x \right)} v_0(s) \mathrm{d} s$ is the modal convolution term, which, being independent of the sample specific functional form of $\taup(x)$, can be pre-computed.
This formulation allows for efficient and precise evaluation of the effective frictional strength $\taucr=\min \taun(x)$ for a large number of samples $N=10,000$, such that the probability distribution $f(\taucr)$ and its evolution as function of the correlation length $\corlen$ can be accurately studied.

\subsection{Monte Carlo Results}

\begin{figure}
\begin{center}
  \includegraphics[width=6.375in]{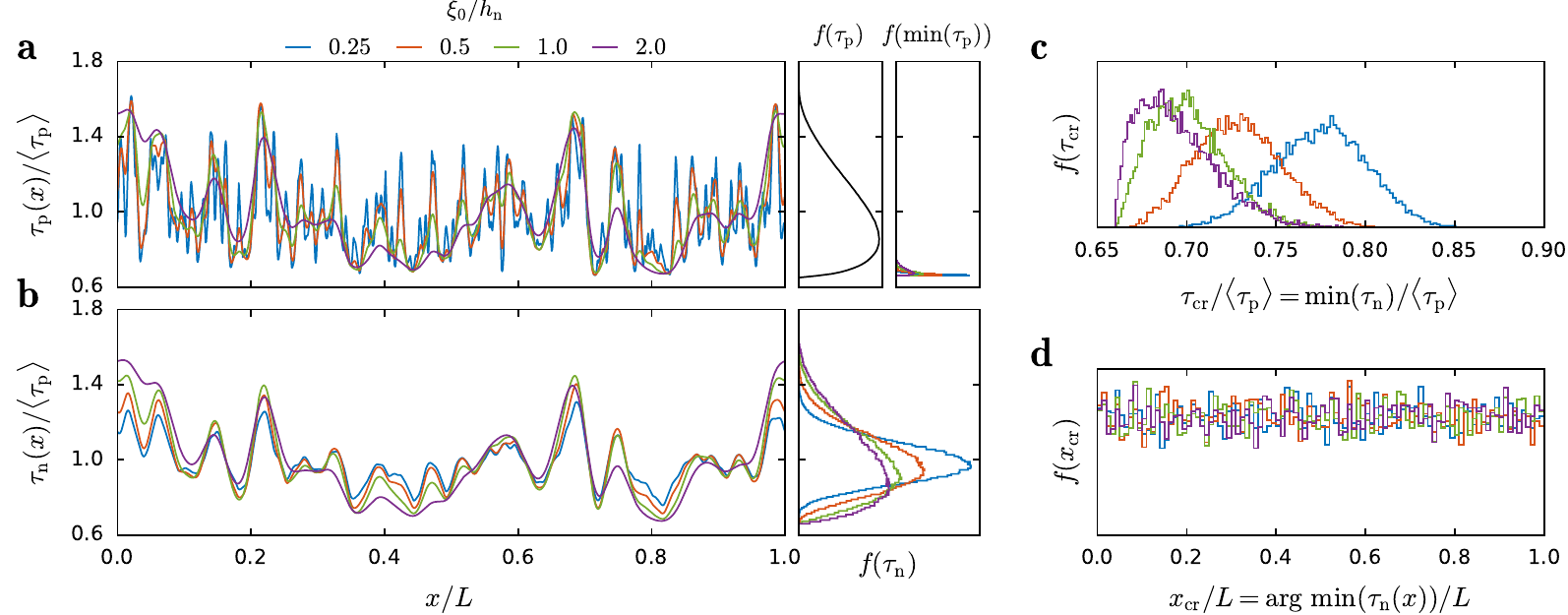}
  \caption{Analytical Monte Carlo study.
  (a)~The random local friction strength $\taup(x)$ is generated with different correlation lengths $\corlen$. For visual purposes, the same random seed is used for the 4 cases shown. 
  (b)~The corresponding local nucleation strength $\taun(x)$ is computed using (Eq.~\ref{eq:taun}). 
  The probability densities $f$ of the random fields $\taun$, $\min(\taun)$ and $\taup$ are reported on the right of (a) and (b), respectively and computed using $N=10,000$ samples. 
  $f(\min(\taup))$ and $f(\taun)$ depend on $\corlen$. 
  %Simplified model for $f(\taun)$, given by Eq.~\ref{eq:simplified_ftaun}, is shown as dashed lines.
  (c) Probability density of the global friction strength $\taucr$. 
  %Simplified model for $f(\taucr)$, given by Eq.~\ref{eq:simplified_ftaucr}, is shown as dashed lines.
  (d) Probability density of the position of the critical nucleation patch $\xnuc$. Note that the seed is not fixed anymore for the samples used in (c) and (d).}
  \label{fig:mc}  
\end{center}
\end{figure}

Prior to presenting the numerical results we provide some intuition of the effect of correlation length  on the nucleation strength $\taun$ based on probabilistic arguments.
By exploiting the stationarity of $\taup$ and $\taun$ it is possible to derive an analytical expression of the expectation of the nucleation strength $\E[\taun]$ and its variance $\Var[\taun]$ as function of the corresponding statistical properties of local strength, $\E[\taup]$, $\Var[\taup]$ and $\corlen/\hn$ (see \ref{sec:appendixStats}).

One interesting finding is that the expectation is not affected by $\corlen/\hn$: $\E[\taun]=\E[\taup]$ (see derivation in Eq.~\ref{eq:AnnexBExp}).
The expression for $\Var[\taun]$, however, involves a double integral of the product of the correlation function $C(.)$ and the eigenfunction $v_0(.)$, which can be evaluated numerically (see derivation in Eq.~\ref{eq:AnnexBVar}).
%For uncorrelated $\taup$, \ie, $\corlen/\hn=0$,  $C(.)$ becomes a Dirac-$\delta$ thus $\Var[\taun]=0$. 
For perfect correlation, \ie, $\corlen/ \hn=\infty$, $C(.)$ becomes a constant, thus $\Var[\taun]=\Var[\taup]$.
Additionally, in the limit of $\corlen\ll \hn$, the double integral in Eq.~\ref{eq:AnnexBVar} scales with $\corlen/\hn$, thus  $\Var[\taun]\propto \Var[\taup]{\corlen}/{\hn}$ (see derivation in Eq.~\ref{eq:ABVarll}).

%Suppose that the integral in Eq.~\ref{eq:taun} is approximated by $\taun= \frac{1}{\hn}\sum_{i=1}^N\taup(\xi_i+x)\Delta\xi_i$, where $\Delta\xi_i=\hn/N$ and $\xi_i=-\hn/2+(i+1/2)\Delta\xi_i$. Then, the mean and variance of $\taun$ become $\E[\taun]=E[\taup]$ and $\Var[\taun]=\frac{\Var[\taup]}{N^2}\sum_{i,j=1}^N C(\xi_i-\xi_j)$; for perfect correlation (\ie, $\corlen=\infty$) $\Var[\taun]=\Var[\taup]$ and for uncorrelated $\taup$ (\ie, $\corlen=0$) $\Var[\taun]=\Var[\taup]/N$. Assume now that $C(\xi)\approx e^{-(\xi/\corlen)^2}$ and let $N\rightarrow \infty$; the variance can be estimated analytically as function of the correlation length\footnote{$\sum_{i,j=1}^N C(\xi_i-\xi_j) = \frac{1}{\hn^2}\iint_{-\hn/2}^{\hn/2}C(s-t)\mathrm{d}s\mathrm{d}t$}  $\Var[\taun]=\Var[\taup]\left(({\corlen}/{\hn})^2\left\(\exp({-(\hn/\corlen)^2})-1\right) + ({\corlen}/{\hn}) \sqrt{\pi}\mathrm{erf}(\hn/\corlen)\right)$.  If  the correlation length is small, $\corlen/\hn<1$ the variance becomes $\Var[\taun]\approx\Var[\taup]\left(({\corlen}/{\hn})^2\sqrt{\pi} -({\corlen}/{\hn})\right)$. However, if the correlation length is large $\corlen/\hn>4$ the variance of the nucleation strength corresponds to the variance of the local strength $\Var[\taun]\approx\Var[\taup]$.

We consider a range of correlation lengths $\corlen/\hn=\{0.25,0.5,1.0,2.0\}$, while all other properties remain constant. 
Fig.~\ref{fig:mc}a-left shows one sample of $\taup(x)$ for each considered $\corlen$.
For clarity of visualization, in Fig.~\ref{fig:mc}a we use the same seed when generating the random fields. Hence, the fields have the same modal random amplitudes $A_j$ and $B_j$, see Eq.~\ref{eq:gaussian_process}, but have different modal spectral densities $\sigma_j^2$, corresponding to the different $\corlen$. 
For this reason, all shown samples have a similar spatial evolution and the effect of varying $\corlen$ can be clearly observed.
By definition, all $\taup(x)$ samples are drawn from the same probability distribution $f(\taup)$ (see Fig.~\ref{fig:mc}a-center).
Decreasing $\corlen$, moves the probability density of its minimum $f(\min(\taup))$ towards the lower bound $\taupmin=0.66\langle \taup\rangle$ (see Fig.~\ref{fig:mc}a-right), because with lower correlation lengths it is more likely to visit a broad range of $\taup$ values.

Fig.~\ref{fig:mc}b-left shows the corresponding nucleation strength $\taun(x)$ for each of the local frictional strength fields $\taup(x)$ presented in Fig.~\ref{fig:mc}a, computed using Eq.~\ref{eq:taun_spec}. 
As mentioned before, $\taun$ is essentially a weighted moving average of $\taup$ with window size $\hn$ (see Eq.~\ref{eq:taun}). 
Thus, most of the high frequency content of $\taup$ disappears and the effect of $\corlen$ on $\taun$ is more subtle.
One interesting feature is in the minima and maxima of $\taun$: increasing $\corlen$ causes lower minima and higher maxima, because the moving average is effectively computed over an approximately constant field $\taun \approx \taup$.
Inversely, decreasing $\corlen$ causes the opposite effect and $\taun\approx\langle \taup\rangle$.

This effect is more clearly visible by considering the distribution $f(\taun)$ shown in Fig.~\ref{fig:mc}b-right.
Increasing $\corlen$ effectively puts more weight onto the tails of $f(\taun)$ (see $\corlen/\hn=2.0$ in Fig.~\ref{fig:mc}b-right), and in the limiting case of $\corlen/\hn\rightarrow \infty$ the distribution of $\taun$ will be the same as the one of $\taup$ (analogous to  Eq.~\ref{eq:ABVarinf}).
On the other hand, decreasing $\corlen$ puts weight on its mean $\langle \taup \rangle$, making $f(\taun)$ similar to a Gaussian (see $\corlen/\hn=0.25$ in Fig.~\ref{fig:mc}b-right) with variance proportional to $\corlen$ (see Eq.~\ref{eq:ABVarll}). In the limit $\corlen/\hn\rightarrow 0$ the distribution of $\taun$ becomes a Dirac-$\delta$ centered at $\langle \taup\rangle$.  
The described dependence of $f(\taun)$ on $\corlen$  confirms the previously stated statistical arguments (see \ref{sec:appendixStats} for derivation).

Because $f(\taup)$ is skewed towards the lower bound of $\taup$ so is $f(\taun)$; the larger $\corlen$ the larger the skewness.
For $\taucr$ this effect is amplified by the fact that $\taucr=\min(\taun(x))$ as depicted in Fig.~\ref{fig:mc}c, causing $\langle \taucr \rangle$ to decrease with increasing $\corlen$. 
As noted in Sec.~\ref{sec:results}, the location where the critical instability occurs $\xnuc$ is uniformly distributed over the entire domain as shown in Fig.~\ref{fig:mc}d and is independent on $\corlen$.

\begin{figure}
\begin{center}
  \includegraphics[width=4.5in]{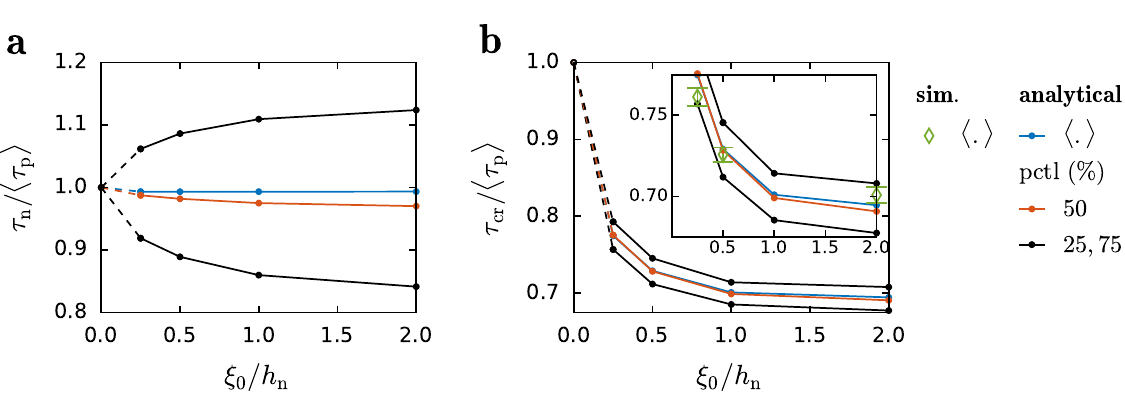}
  \caption{Variation of nucleation strength $\taun$ (a) and effective friction strength $\taucr$ (b) as function of correlation length $\corlen$. 
  Solid lines are the results from the analytical Monte Carlo study with $N=10,000$ (same data as Fig.~\ref{fig:mc}).
  Data-points for $\corlen/\hn\rightarrow 0$ are based on analytical considerations and connected to the analytical Monte Carlo results by dashed lines.
  Diamonds are results from 60 dynamic simulations (same data as Fig.~\ref{fig:theory_vs_sim}).}
  \label{fig:mc_vs_xi}  
\end{center}
\end{figure}

We further analyze the effect of $\corlen$ on the probability distribution of $\taun$ and $\taucr$ by reporting the mean, median and $25\%$ percentile of the probability density function (see Fig.~\ref{fig:mc_vs_xi}). We observe that the nucleation strength tends towards the mean peak strength for vanishing correlation length, $\lim_{\corlen/\hn \rightarrow 0} \taun(x)=\langle \taup\rangle$, because the moving average in computing $\taun(x)$ is evaluated over a window $\hn$ that appears infinite compared to $\corlen$ (see Fig.~\ref{fig:mc_vs_xi}a). Consequently, the effective strength also tends towards the mean of the local strength: $\lim_{\corlen/\hn \rightarrow 0} \taucr=\langle \taup\rangle$ (see Fig.~\ref{fig:mc_vs_xi}b).
Conversely, if ${\corlen \gg \hn}$, the moving average is computed over a window $\hn$ which vanishes, and thus (\ref{eq:taun}) becomes the identity:  $\lim_{\corlen/\hn \rightarrow \infty}\taun(x) = \taup(x)$.
In this case, the effective strength will be more likely to be close to the actual lower bound of the distribution $\lim_{\corlen/\hn \rightarrow \infty}\taucr = \min(\taup)$ (see Fig.~\ref{fig:mc_vs_xi}b).
The transition between these two limiting cases is described by the results of the analytical Monte Carlo study, which are validated by 20 dynamic simulations for each $\corlen/\hn$ by reporting the mean effective friction strength (see inset in Fig.~\ref{fig:mc_vs_xi}b). 
For $\corlen/\hn\geq0.5$ simulations and theory coincide. However, for $\corlen/\hn=0.25$ the theoretical model slightly overestimates the effective friction strength. In Sec.~\ref{sec:discussion}, we will argue that the lower effective friction at small $\corlen/\hn$ is caused by interactions between neighbouring nucleation patches, which destabilize each other and lead to unstable growth at lower overall loading compared to the theoretical prediction.

\section{Discussion}\label{sec:discussion}

\subsection{Implications of the Physical Problem}\label{sec:implphys}

The analyzed physical problem is simplistic and contains only the absolute minimum of a realistic system with a frictional interface -- while still maintaining a rigorous representation of the constitutive relation of the bulk and the interface.
The objective is to provide a fundamental understanding of the macroscopic effects on static friction caused by randomness in the local frictional properties. While many options exist to complexify the proposed system, we leave them for future work and focus here on the basics. Nevertheless, in this section, we will discuss some of these simplifications as well as their implications.

Randomness along the interface may have various origins including heterogeneity in bulk material properties and local environmental conditions (\eg, humidity and impurities). Prominent causes for randomness are geometric imperfections, which include non-flat interfaces and surface roughness. The real contact area, which is an ensemble of discrete micro-contacts \citep{bowden_friction_1950,dieterich_direct_1994,li_micromechanics_2008,sahli_evolution_2018} and is much smaller than the apparent contact area, introduces naturally randomness to the interface. Surface roughness is often modeled as self-affine fractals \citep{pei_finite_2005}, which directly affects the size distribution of micro-contacts and local contact pressure. The resulting frictional properties are expected to vary similarly. This would typically lead to small areas of the interface with high frictional strength and most areas with no resistance against sliding, \ie, $\taup = 0$, since only the micro-contacts may transmit stresses across the interface. Therefore, at this length scale, one would expect the random strength field to be bound by zero at most locations, similar to the approach taken by \citet{barras_onset_2019}. However, in many engineering systems, the nucleation length is orders of magnitude larger than the characteristic length scales of the micro-contacts: nucleation lengths of $\sim 10-100~\mathrm{mm}$ \citep{ben-david_static_2011,latour_characterization_2013} and surface roughness lengths of $\sim 1~\mu\mathrm{m}$ \citep{svetlizky_classical_2014}. For this reason, we consider a continuum description with a somewhat larger length scale. In our approach, the frictional strength profile is continuous and varies due to randomness in the micro-contacts population without considering individual contact points.

% check Garagash and Germanovich for discussion 7.1 of rate and state friction 
Surface roughness and other local properties directly affect how frictional strength changes depending on slip $\slip$, slip rate $\sliprate$, and state \citep{rabinowicz_friction_1995,pilvelait_influences_2020}. This is often modeled in phenomenological rate-and-state friction models \citep{dieterich_modeling_1979,ruina_slip_1983,rice_stability_1983}. 
As discussed by \citet{garagash_nucleation_2012} and demonstrated by \citet{rubin_earthquake_2005} and \citet{ampuero_earthquake_2008}, the nucleation length scale of rate-and-state friction models approaches asymptotically the critical length $\hn$ used in this work and given by Eq.~\ref{eq:critlength} if the rate-and-state friction parameters are favoring strong weakening with slip rate. However, if rate-weakening becomes negligible, the nucleation criterion tends towards the Griffith's length \citep{andrews_rupture_1976}, which applies to ruptures with small-scale yielding. In this case, the frictional weakening process is contained in a small zone at the rupture tip and most of the rupture surface is at the residual stress level, which is different to the nucleation patches by \citet{uenishi_universal_2003}, where the entire rupture surface is still weakening when the critical length is reached. 

Since many engineering materials present relatively important slip-rate weakening friction, \eg, dynamic weakening of $\sim 1~\mathrm{MPa}$ for glassy polymers at normal pressure of $\sim 5~\mathrm{MPa}$ \citep{svetlizky_dynamic_2020}  and, similarly, $\sim 1~\mathrm{MPa}$ weakening for granite at normal pressure of $\sim 6~\mathrm{MPa}$ \citep{kammer_fracture_2019}, we considered a model system with strong frictional weakening. However, we neglect the complexity of rate-and-state friction, as extensively demonstrated by \citet{ray_earthquake_2017,ray_homogenization_2019}, and apply a linear slip-weakening friction law at the interface because it has the most important features of friction, \ie, a weakening mechanism, while being simple and well-understood. The advantage is that the weakening-rate $\weakrate$ is predefined. It further has a well-defined \emph{fracture} energy $\fracE$, which is the energy dissipated by the weakening process, \ie, the triangular area $(\taup - \taur)\dc/2$ in Fig.~\ref{fig:model}c:
\begin{equation}
    \fracE(x) = \frac{(\taup(x) - \taur)^2}{2\weakrate} ~.
    \label{eq:fracenergy}
\end{equation}
Since $\weakrate$ is constant in our system $\fracE$ varies with $(\taup(x)-\taur)^2$. A correlation between $\fracE(x)$ and $\taup(x)$ can be expected given that any point that is stronger, \ie, increased $\taup$, is likely to dissipate more energy as well, \ie, increased $\fracE$.

Further, the linear slip-weakening law is contact-pressure independent, which may appear counter-intuitive based on Coulomb's well-known friction laws \citep{amontons_resistance_1699,coulomb_theorie_1785}. However, the contact pressure is, due to symmetry in similar-material interfaces, constant over time and, therefore, any possible pressure dependence becomes irrelevant for the nucleation process itself. Nevertheless, local friction properties are expected to change for systems with different normal pressure. This effect has not been analyzed here since we did not vary the contact pressure, but could be taken into account by changing the values of $\taup$, $\taur$ and $\dc$.

Finally, we note that by assuming a periodic system, we neglect possible boundary effects. We expect that the boundary would locally reduce $\taucr$ compared to the prediction based on Eq.~\ref{eq:taucr}, which assumes an infinite domain, because the free boundary would locally restrict stress redistribution and thus increase the stress at the edge of the nucleation patch. Therefore, the probability density of global frictional strength $\taucr$ for a periodic system, as shown in Fig.~\ref{fig:mc}c, has likely a slight tendency towards higher values compared to a finite system. However, we expect the spatial range of the boundary effect to scale with $\hn$ and, therefore, $f(\taucr)$ will tend towards the periodic solution for $\hn/\sysL \rightarrow 0$. Verification would require a large number of numerical simulations, which is beyond the scope of this work.

% DSK: in my opinion, we do not need this.
%Further simplification of the studied friction problem is still possible. This would involve replacing the elastic response of the infinite half space with a Burridge-Knopoff-type model, which considers an array of sliders coupled to each-other and to a rigid interface by linear springs \citep{burridge_model_1967}. Variations of the Burridge-Knopoff model have been extensively used to  study frictional systems due to the ease of deriving analytical solutions and stability analysis \citep{carlson_properties_1989,amon_avalanche_2017}. However, such simple models only provide a crude approximation of the long range interactions characteristic of the bulk's elastic response, which means that are more suitable for qualitative studies.

\subsection{Interpretation of Numerical Simulations}
\label{sec:interpretsimulation}

The simulations have shown that uniform $\taui$ and random $\taup(x)$ cause multiple nucleation patches to develop simultaneously. We can see in Fig.~\ref{fig:simulation}b-right that $20$-$30$ patches (bright blue stripes) coexist by the time global strength is reached, \ie, $t/T = 1$. Most of these nucleation patches grow very slowly and their number increases with increasing $\taui(t)$. Nucleation patches can also merge, which is what happens in this simulation to the critical patch. Furthermore, the simulation shows that unstable growth and thus global failure is not necessarily caused by the first nucleation patch to appear. For instance, the $\taup(x)$ profile shown in Fig.~\ref{fig:simulation}c presents three local minima with approximately the same value, \ie, at $x/\sysL \approx 0.4$, $0.6$ and $0.75$. Therefore, the first three nucleation patches appear quasi-simultaneously. Whether one of these patches or another one appearing later is the one becoming unstable first does only dependent indirectly on the minimum value of $\taup(x)$. More important is whether $\taup(x)$ remains low in the near region of the local minimum. The nucleation patch at $x/\sysL \approx 0.75$ is in an area of relatively low $\taup(x)$, compared to the other early nucleation patches, which is why it develops faster to the critical size and causes unstable propagation.

This non-local character of the nucleation patches becomes obvious when considering the integral form of Eq.~\ref{eq:taun} that corresponds to a weighted moving average of $\taup(x)$. In Fig.~\ref{fig:theory_vs_sim}a, we can see that $\taun$ at $x/\sysL \approx 0.75$ is considerably lower than at the location of the other early nucleation patches $x/\sysL \approx 0.4$ and 0.6. This is why $x/\sysL \approx 0.75$ gets critical first and causes unstable slip area growth. Interestingly, $x/\sysL \approx 0.3$ is the second most critical point even though the local minimum in $\taup$ is higher than many others in this system. However, $\taup(x)$ remains rather low over an area that approaches $\hn$, and therefore $\taun$ is also low.

In Fig.~\ref{fig:theory_vs_sim}, we compared the prediction of $\taucr$ with measurements from simulations and showed that the prediction works generally well. However, we noticed that for decreasing $\corlen/\hn$ the discrepancies increase. The theory generally predicts higher $\taucr$ than observed in simulations. We believe that this is caused by nucleation patch interaction, which is neglected in the current theory. The interaction may occur if two nucleation patches are near each other. Nucleation patches cause stress redistribution because the stress inside the patch decreases but, due to equilibrium, it increases in the area near the patch. Therefore, a nucleation patch may cause an increase of the effective stress (compared to the applied load) to another patch, which leads to increased patch size and thus $\hn$ is reached already at $\taucrsim < \taucrpred$. Interestingly, a similar phenomenon has recently been observed in simulations of compressive failure governed by a mesoscopic Mohr-Coulomb criterion, where local damage clusters interact and eventually coalesce to macroscopic failure \citep{dansereau_collective_2019}. 

The nucleation patch interaction is likely also the cause for discrepancies observed in the prediction of the nucleation location $\xnuc$, as shown in Fig.~\ref{fig:theory_vs_sim}c. While most cases are very well predicted, some simulations present unstable growth that starts from a different location. In these cases, two nucleation patches have very similar critical stress level. However, the (slightly) less critical patch interacts with a neighboring smaller patch and thus becomes unstable at a lower stress level than theoretically expected. This is more likely to occur for systems with low $\corlen/\hn$ since this increases the likelihood of another local minimum being located close to active nucleation patches. 
Nevertheless, the critical stress level $\taucr$ remains quantitatively well predicted, as shown Fig.~\ref{fig:theory_vs_sim}b and discussed above, because these secondary effects are minor.

The representative simulation illustrated in Fig.~\ref{fig:simulation} shows that the frictional rupture front, after becoming unstable, does not arrest until it propagated across the entire interface leading to global sliding. This is a general feature of our problem and all our simulations present the same behavior. What is the reason for this run-away propagation? 
%Initially, the nucleation patch continues to weaken along its entire length, \ie, $\slip < \dc$ everywhere. 
Right after nucleation, the slipping area continues to weaken along its entire length, \ie, $\slip < \dc$ everywhere. 
However, after some more growth, it transforms slowly into a frictional rupture front, which is essentially a Griffith's shear crack with a cohesive zone and constant residual strength~\citep{svetlizky_classical_2014,svetlizky_dynamic_2020,garagash_nucleation_2012}. The arrest of frictional rupture fronts are governed by an energy-rate balance \citep{kammer_linear_2015}, which states that a rupture continues to propagate as long as the (mode II) static energy release rate $\staticERR$ is larger than the fracture energy, \ie, $\staticERR > \fracE$. In our system, the stress drop $\Dtau = \taui - \taur$ is uniform since $\taui$ and $\taur$ are uniform. Thus, the static energy release rate grows linearly with rupture length $\staticERR \propto \cracklength$, and it becomes increasingly difficult to arrest a rupture as it continues to grow. Specifically for our case, we find that $\staticERR > \fracE_\mathrm{max}$ for $\cracklength/\hn \gtrapprox 3$, where $\fracE_\mathrm{max}$ is $\fracE$ from Eq.~\ref{eq:fracenergy} for $\taupmax$. Hence, once the slipping area reached a size of $\cracklength/\hn \gtrapprox 3$, nothing can stop it anymore -- not even $\taupmax$.
For $\cracklength/\hn  \lessapprox 3$, it is theoretically possible for the slipping area to arrest after some unstable propagation. However, a large increase in $\taup(x)$ would need to occur simultaneously on both side, which is very unlikely, in particular for $\corlen/\hn > 1$. 
%Determining the precise probability for potential re-arrest is beyond the scope of this work and is left for future efforts.
Therefore, our assumptions of constant stress drop $\Dtau$ and limited variation of local frictional strength causes arrested rupture fronts to be extremely rare. 
Nevertheless, since $\fracE_\mathrm{max}$ depends on the probability distribution function $f(\taup)$, a larger variance, and thus a larger $\taupmax$, would make crack arrest (slightly) more likely.

It is interesting to note, however, that arrest of dynamically propagating slipping areas may occur in other systems. \citet{amon_avalanche_2017}, for instance, showed in their simulations that multiple smaller events nucleate and arrest in order to prepare the interface for a global event. In their system, the initial position along the interface is random as well as the friction properties.
Therefore, the available elastic energy, which is the driving force, is random and may fluctuate enough to cause arrest. For the same reason, \citet{geus_how_2019} observed arrested events of various sizes in simulations with random potential energy along the interface. On the contrary, our system, as outlined above, is characterized by steadily increasing available energy and, thus, behaves differently.  

Experimental evidence for arrest of frictional rupture fronts is rather limited. The arrest of confined events observed by \citet{rubinstein_dynamics_2007} on glassy polymers and by \citet{ke_rupture_2018} on granite, is caused by non-uniform loading due to the experimental configuration as demonstrated by \citet{kammer_linear_2015} and \citet{ke_rupture_2018,ke_analytical_2019}. While small scale randomness in the applied shear stress may occur, it does not cause arrest -- at best, it may slightly delay or expedite it. Therefore, these experimental observations do not support the presence of any important randomness in the applied shear load; at least at these scales. In much larger systems, such as tectonic plates, randomness in the background stress is likely very important, as discussed in Sec.~\ref{sec:earthquakes}.

%we have $\taup/\langle \taup \rangle \in (0.66,1.66)$.
%What would happen if $\taup/\langle \taup \rangle \in (0,10)$

\subsection{Interpretation of Monte Carlo Study}\label{sec:impl_mc}

In an engineering context, it is usually not enough to know the mean value of a macroscopic property, \eg, the static friction strength, since design criteria are determined based on probability of failure; and risk assessments require failure probability analysis. 
If the stochastic properties of local interfacial strength $\taup(x)$ are known, the developed theoretical framework in Sec.~\ref{sec:analyicalMCstudy} provides a tool to evaluate the global strength distribution and, hence, the failure probability.
However, $\taup(x)$ is not directly observable in experiments (at least so far). 
In the absence of experimental evidence, the stochastic properties of $\taup(x)$ have been chosen based on physical considerations but our specific parameter choice is arbitrary. 
This approach is a first approximation that enables us to develop an efficient Monte Carlo method to study the effects of such variations on macroscopic properties. 
This method may be adapted to more realistic random friction profiles.
In the following section we will discuss the choice of each stochastic property of $\taup(x)$ and its effect on the variability of global strength, $\taucr$.

We assumed that $\taup(x)$ is a random variable following a Beta distribution because it provides simultaneously a non-Gaussian property and well-defined boundaries for minimum and maximum strength, which is physically consistent since mechanical properties are bounded. 
The parameters of the Beta distribution are chosen such that it is skewed towards the lower bound of $\taup$. 
Physically this means that the local interface strength is mostly weak with few strong regions.
Under this assumption we observed that the global interface strength $\taucr$ is close the  lower bound of $\taup$ and that decreasing correlation length $\corlen$ leads to higher $\taucr$ with smaller variation (see Fig.~\ref{fig:mc_vs_xi}).
Nucleation is governed by the strength of the weakest region which size equals or exceeds the nucleation length.
Hence, if the skewness would be toward high values of $\taup$ -- this would corresponds to a mostly strong interface with few weak regions -- the variation of $\taucr$ would be larger because of the longer tail at the lower bound.
However, the effect of correlation length would remain unchanged: lower $\corlen$ would cause higher $\taucr$ with smaller variation.

Further, we assumed that $\taup$ has a power spectral density specified in Eq.~\ref{eq:markov4} with a specific exponent, which affects the memory of the random field.
A smaller exponent would result in a flatter decay above the cutoff frequency, and thus generate a field with more high-frequency content. %, which, due to the moving average effect of the convolution, would result in higher f (τn ) in the proximity of the mean and, consequently, slightly higher τcr .
However, when considering equivalent correlation lengths\footnote{$C(\xi_{0})=e^{-1}$}, effects of different assumptions regarding the functional form of the power spectral density are expected to be minor in the probability density of $\taucr$.

It is noteworthy that there are three relevant length scales: $\sysL$, $\hn$ and $\corlen$. Here, we have not considered the effects of changes in $\sysL$ so far. Based on our theoretical model, we expect that a larger $\sysL$ would result in smaller variance of $\taucr$. One approach to explore this effect, while avoiding to change the size of the experimental system, could be to modify the normal load, which would affect $\taup$ and $\taur$, and hence the critical nucleation size $\hn \propto (\taup-\taur)^{-1}$, as shown experimentally by \citet{latour_characterization_2013}.

\subsection{Implications for Earthquake Nucleation}\label{sec:earthquakes}

In the current study, we are interested in estimating the probability distribution of the macroscopic strength of a frictional interface of given size $\sysL$. Similar systems but with a focus on other aspects have been studied in order to gain a better understanding of earthquake nucleation. 
The challenges in studying earthquake nucleation are associated with the size of the system (hundreds of kilometers) and the limited physical access to measure important properties, such as stress state and frictional properties. 
However, when and how an earthquake nucleates affects directly the average stress drop level $\langle \Dtau \rangle \equiv \taucr - \taukin$, and the earthquake magnitude, which is more easily determined. Hence, there is a need to infer from earthquake magnitude observation back on the fault properties and their variability, to learn about the risk of potential future earthquakes. This is a similar inverse problem as described above.  

Previous studies have shown that randomness in simplified models present earthquake magnitudes that follow a power-law distribution \citep{carlson_properties_1989,ampuero_properties_2006}. The simulations presented a large range of magnitudes because the slipping areas arrested, which is the result of randomness in the local stress drop, as discussed in Sec.~\ref{sec:interpretsimulation}. Interestingly, small events were shown to smoothen the stress profile, which reduces the randomness, thus prepared the interface for larger events, as also observed experimentally \citep{ke_rupture_2018}. This would suggest that nucleation of larger events tend to be caused by randomness in fault properties rather than (background) stress level, since the stress is getting smoothed. Therefore, our model provides a simple but reasonable tool to study nucleation of medium to large earthquakes.

Our results show that smaller correlation length lead to higher overall strength and more variation. In the context of earthquakes, this would suggest that smaller $\corlen/\hn$ support larger earthquake magnitudes since nucleation at a higher stress level translates into larger average stress drops, which provides more available energy to release and, thus, makes arrest more difficult. This is complementary to observations by \citet{ampuero_properties_2006} that showed a trend to higher earthquake magnitudes for decreasing standard deviation of the random stress drop field $\Dtau(x)$ while keeping $\corlen/\hn$ constant.  

In addition to the important question on critical stress level for earthquake nucleation, it also remains unclear how the nucleation process takes place. Two possible models \citep{beroza_properties_1996,noda_large_2013,mclaskey_earthquake_2019} are discussed. 
The "Cascade" model, where foreshocks trigger each other with increasing size and finally lead to the main earthquake; and the "Pre-Slip" model, which assumes that nucleation is the result of aseismic slow-slip.
In our simplistic model, nucleation occurs in a "Pre-Slip" type process, with a long phase of slow slip and an abrupt acceleration after the nucleation patch reached its critical size (see Fig.~\ref{fig:simulation}b-right). However, if the amplitude range of our random $\taup(x)$ field was much larger, the likelihood of arrest would increase and, thus, interaction between arrested small events could emerge. This would lead to a nucleation process that resembles more the "Cascade" model. We, therefore, conclude that the type of nucleation process that may occur at a given fault depends on extent of randomness in the local stress and property fields.

\section{Conclusion}
\label{sec:conclusion}

We studied the stochastic properties of frictional interfaces considering the nucleation of unstable slip patches. 
We considered a uniform loading condition and studied the effect of random interface strength, characterized by its probability density and correlation function. 
Using numerical simulations solving the elastodynamic equations, we demonstrated that macroscopic sliding does not necessarily occur when the weakest point along the interface starts sliding, but when one of possible many slowly slipping nucleation patches reaches a critical length and becomes unstable. 
We verified that the nucleation criterion originally developed by \citet{uenishi_universal_2003} predicts well the critical stress leading to global sliding if the criterion is formulated as a minimum of the local strength convolved with the first eigenfunction of the elastic problem. 
The simulations further showed that increasing correlation lengths of the random interface strength lead to reduced macroscopic static friction.
Using the theoretical nucleation criterion, we perform a Monte Carlo study that provided an accurate description of the underlying probability density functions for these observed variations in macroscopic friction. 
We showed that the probability density function of the global critical strength approaches the probability density of the minimum in the random local strength when the correlation length is much larger than the critical nucleation length. Conversely, a vanishingly small correlation length results in generally higher macroscopic strength with smaller variation.
We showed that the presence of precursory dynamic slip events, as in more complex models, is extremely unlikely under the assumption of uniform stress drop.
Finally, we discussed discrepancies between the theoretical model and simulations, which suggest that for small correlation lengths the theoretical prediction overestimates the frictional strength, possibly because it neglects interactions between neighboring nucleation patches. 
%Finally, we provided a simplified analytical model expressing the probability distribution of the global interface strength as function of the stochastic properties of the local strength, which supplements the Monte Carlo study.

% ------------------------------------------------------------------------------------
%\section*{References}
\bibliography{p22-stoch-fric}

\begin{thebibliography}{58}
\expandafter\ifx\csname natexlab\endcsname\relax\def\natexlab#1{#1}\fi
\providecommand{\url}[1]{\texttt{#1}}
\providecommand{\href}[2]{#2}
\providecommand{\path}[1]{#1}
\providecommand{\DOIprefix}{doi:}
\providecommand{\ArXivprefix}{arXiv:}
\providecommand{\URLprefix}{URL: }
\providecommand{\Pubmedprefix}{pmid:}
\providecommand{\doi}[1]{\href{http://dx.doi.org/#1}{\path{#1}}}
\providecommand{\Pubmed}[1]{\href{pmid:#1}{\path{#1}}}
\providecommand{\bibinfo}[2]{#2}
\ifx\xfnm\relax \def\xfnm[#1]{\unskip,\space#1}\fi
%Type = Article
\bibitem[{Amon et~al.(2017)Amon, Blanc and Géminard}]{amon_avalanche_2017}
\bibinfo{author}{Amon, A.}, \bibinfo{author}{Blanc, B.},
  \bibinfo{author}{Géminard, J.C.}, \bibinfo{year}{2017}.
\newblock \bibinfo{title}{Avalanche precursors in a frictional model}.
\newblock \bibinfo{journal}{Physical Review E} \bibinfo{volume}{96},
  \bibinfo{pages}{033004}.
\newblock \URLprefix \url{https://link.aps.org/doi/10.1103/PhysRevE.96.033004},
  \DOIprefix\doi{10.1103/PhysRevE.96.033004}.
%Type = Article
\bibitem[{Amontons(1699)}]{amontons_resistance_1699}
\bibinfo{author}{Amontons, G.}, \bibinfo{year}{1699}.
\newblock \bibinfo{title}{De la résistance causée dans les machines}.
\newblock \bibinfo{journal}{Memoires de l’Academie Royale A} ,
  \bibinfo{pages}{275--282}.
%Type = Incollection
\bibitem[{Ampuero et~al.(2006)Ampuero, Ripperger and
  Mai}]{ampuero_properties_2006}
\bibinfo{author}{Ampuero, J.P.}, \bibinfo{author}{Ripperger, J.},
  \bibinfo{author}{Mai, P.M.}, \bibinfo{year}{2006}.
\newblock \bibinfo{title}{Properties of {Dynamic} {Earthquake} {Ruptures}
  {With} {Heterogeneous} {Stress} {Drop}}, in: \bibinfo{booktitle}{Earthquakes:
  {Radiated} {Energy} and the {Physics} of {Faulting}}.
  \bibinfo{publisher}{American Geophysical Union},
  \bibinfo{address}{Washington, DC}, pp. \bibinfo{pages}{255--261}.
\newblock \URLprefix
  \url{https://resolver.caltech.edu/CaltechAUTHORS:20120830-075047141}.
%Type = Article
\bibitem[{Ampuero and Rubin(2008)}]{ampuero_earthquake_2008}
\bibinfo{author}{Ampuero, J.P.}, \bibinfo{author}{Rubin, A.M.},
  \bibinfo{year}{2008}.
\newblock \bibinfo{title}{Earthquake nucleation on rate and state faults –
  {Aging} and slip laws}.
\newblock \bibinfo{journal}{Journal of Geophysical Research: Solid Earth}
  \bibinfo{volume}{113}.
\newblock \URLprefix
  \url{https://agupubs.onlinelibrary.wiley.com/doi/abs/10.1029/2007JB005082},
  \DOIprefix\doi{10.1029/2007JB005082}.
%Type = Article
\bibitem[{Andrews(1976)}]{andrews_rupture_1976}
\bibinfo{author}{Andrews, D.J.}, \bibinfo{year}{1976}.
\newblock \bibinfo{title}{Rupture velocity of plane strain shear cracks}.
\newblock \bibinfo{journal}{Journal of Geophysical Research (1896-1977)}
  \bibinfo{volume}{81}, \bibinfo{pages}{5679--5687}.
\newblock \URLprefix
  \url{https://agupubs.onlinelibrary.wiley.com/doi/abs/10.1029/JB081i032p05679},
  \DOIprefix\doi{10.1029/JB081i032p05679}.
%Type = Article
\bibitem[{Barras et~al.(2019)Barras, Aghababaei and
  Molinari}]{barras_onset_2019}
\bibinfo{author}{Barras, F.}, \bibinfo{author}{Aghababaei, R.},
  \bibinfo{author}{Molinari, J.F.}, \bibinfo{year}{2019}.
\newblock \bibinfo{title}{Onset of sliding across scales: {How} the contact
  topography impacts frictional strength}.
\newblock \bibinfo{journal}{arXiv:1910.11280 [cond-mat, physics:physics]}
  \URLprefix \url{http://arxiv.org/abs/1910.11280}. \bibinfo{note}{arXiv:
  1910.11280}.
%Type = Article
\bibitem[{Bayart et~al.(2016)Bayart, Svetlizky and
  Fineberg}]{bayart_fracture_2016}
\bibinfo{author}{Bayart, E.}, \bibinfo{author}{Svetlizky, I.},
  \bibinfo{author}{Fineberg, J.}, \bibinfo{year}{2016}.
\newblock \bibinfo{title}{Fracture mechanics determine the lengths of interface
  ruptures that mediate frictional motion}.
\newblock \bibinfo{journal}{Nature Physics} \bibinfo{volume}{12},
  \bibinfo{pages}{166--170}.
\newblock \URLprefix
  \url{http://www.nature.com/nphys/journal/v12/n2/abs/nphys3539.html},
  \DOIprefix\doi{10.1038/nphys3539}.
%Type = Article
\bibitem[{Ben-David and Fineberg(2011)}]{ben-david_static_2011}
\bibinfo{author}{Ben-David, O.}, \bibinfo{author}{Fineberg, J.},
  \bibinfo{year}{2011}.
\newblock \bibinfo{title}{Static {Friction} {Coefficient} {Is} {Not} a
  {Material} {Constant}}.
\newblock \bibinfo{journal}{Physical Review Letters} \bibinfo{volume}{106},
  \bibinfo{pages}{254301}.
\newblock \URLprefix
  \url{https://link.aps.org/doi/10.1103/PhysRevLett.106.254301},
  \DOIprefix\doi{10.1103/PhysRevLett.106.254301}.
%Type = Article
\bibitem[{Beroza and Ellsworth(1996)}]{beroza_properties_1996}
\bibinfo{author}{Beroza, G.C.}, \bibinfo{author}{Ellsworth, W.L.},
  \bibinfo{year}{1996}.
\newblock \bibinfo{title}{Properties of the seismic nucleation phase}.
\newblock \bibinfo{journal}{Tectonophysics} \bibinfo{volume}{261},
  \bibinfo{pages}{209--227}.
\newblock \URLprefix
  \url{http://www.sciencedirect.com/science/article/pii/0040195196000674},
  \DOIprefix\doi{10.1016/0040-1951(96)00067-4}.
%Type = Incollection
\bibitem[{Bilby and Eshelby(1968)}]{bilby_disclocations_1968}
\bibinfo{author}{Bilby, B.}, \bibinfo{author}{Eshelby, J.},
  \bibinfo{year}{1968}.
\newblock \bibinfo{title}{Disclocations and theory of fracture}, in:
  \bibinfo{editor}{Liebowitz, H.} (Ed.), \bibinfo{booktitle}{Fracture, {An}
  {Advanced} {Treatise}}. \bibinfo{publisher}{Academic}, \bibinfo{address}{San
  Diego}. volume~\bibinfo{volume}{1}, pp. \bibinfo{pages}{99--182}.
%Type = Book
\bibitem[{Bowden and Tabor(1950)}]{bowden_friction_1950}
\bibinfo{author}{Bowden, F.}, \bibinfo{author}{Tabor, D.},
  \bibinfo{year}{1950}.
\newblock \bibinfo{title}{The {Friction} and {Lubrication} of {Solids}}.
\newblock \bibinfo{publisher}{Clarendon Press}, \bibinfo{address}{Oxford, UK}.
%Type = Article
\bibitem[{Breitenfeld and Geubelle(1998)}]{breitenfeld_numerical_1998}
\bibinfo{author}{Breitenfeld, M.S.}, \bibinfo{author}{Geubelle, P.H.},
  \bibinfo{year}{1998}.
\newblock \bibinfo{title}{Numerical analysis of dynamic debonding under {2D}
  in-plane and {3D} loading}.
\newblock \bibinfo{journal}{International Journal of Fracture}
  \bibinfo{volume}{93}, \bibinfo{pages}{13--38}.
\newblock \URLprefix
  \url{https://link.springer.com/article/10.1023/A:1007535703095},
  \DOIprefix\doi{10.1023/A:1007535703095}.
%Type = Article
\bibitem[{Campillo and Ionescu(1997)}]{campillo_initiation_1997}
\bibinfo{author}{Campillo, M.}, \bibinfo{author}{Ionescu, I.R.},
  \bibinfo{year}{1997}.
\newblock \bibinfo{title}{Initiation of antiplane shear instability under slip
  dependent friction}.
\newblock \bibinfo{journal}{Journal of Geophysical Research: Solid Earth}
  \bibinfo{volume}{102}, \bibinfo{pages}{20363--20371}.
\newblock \URLprefix
  \url{https://agupubs.onlinelibrary.wiley.com/doi/abs/10.1029/97JB01508},
  \DOIprefix\doi{10.1029/97JB01508}.
%Type = Article
\bibitem[{Carlson and Langer(1989)}]{carlson_properties_1989}
\bibinfo{author}{Carlson, J.M.}, \bibinfo{author}{Langer, J.S.},
  \bibinfo{year}{1989}.
\newblock \bibinfo{title}{Properties of earthquakes generated by fault
  dynamics}.
\newblock \bibinfo{journal}{Physical Review Letters} \bibinfo{volume}{62},
  \bibinfo{pages}{2632--2635}.
\newblock \URLprefix
  \url{https://link.aps.org/doi/10.1103/PhysRevLett.62.2632},
  \DOIprefix\doi{10.1103/PhysRevLett.62.2632}.
%Type = Article
\bibitem[{Coulomb(1785)}]{coulomb_theorie_1785}
\bibinfo{author}{Coulomb, C.}, \bibinfo{year}{1785}.
\newblock \bibinfo{title}{Théorie des machines simples, en ayant regard au
  frottement de leurs parties, et à la roideur des cordages}.
\newblock \bibinfo{journal}{Memoires deMathematique et de Physique de
  l’Academie Royale} , \bibinfo{pages}{161--342}.
%Type = Article
\bibitem[{Dansereau et~al.(2019)Dansereau, Démery, Berthier, Weiss and
  Ponson}]{dansereau_collective_2019}
\bibinfo{author}{Dansereau, V.}, \bibinfo{author}{Démery, V.},
  \bibinfo{author}{Berthier, E.}, \bibinfo{author}{Weiss, J.},
  \bibinfo{author}{Ponson, L.}, \bibinfo{year}{2019}.
\newblock \bibinfo{title}{Collective {Damage} {Growth} {Controls} {Fault}
  {Orientation} in {Quasibrittle} {Compressive} {Failure}}.
\newblock \bibinfo{journal}{Physical Review Letters} \bibinfo{volume}{122},
  \bibinfo{pages}{085501}.
\newblock \URLprefix
  \url{https://link.aps.org/doi/10.1103/PhysRevLett.122.085501},
  \DOIprefix\doi{10.1103/PhysRevLett.122.085501}.
%Type = Article
\bibitem[{Dieterich(1979)}]{dieterich_modeling_1979}
\bibinfo{author}{Dieterich, J.H.}, \bibinfo{year}{1979}.
\newblock \bibinfo{title}{Modeling of rock friction: 1. {Experimental} results
  and constitutive equations}.
\newblock \bibinfo{journal}{Journal of Geophysical Research: Solid Earth}
  \bibinfo{volume}{84}, \bibinfo{pages}{2161--2168}.
\newblock \URLprefix
  \url{https://agupubs.onlinelibrary.wiley.com/doi/abs/10.1029/JB084iB05p02161},
  \DOIprefix\doi{10.1029/JB084iB05p02161}.
%Type = Article
\bibitem[{Dieterich and Kilgore(1994)}]{dieterich_direct_1994}
\bibinfo{author}{Dieterich, J.H.}, \bibinfo{author}{Kilgore, B.D.},
  \bibinfo{year}{1994}.
\newblock \bibinfo{title}{Direct observation of frictional contacts: {New}
  insights for state-dependent properties}.
\newblock \bibinfo{journal}{pure and applied geophysics} \bibinfo{volume}{143},
  \bibinfo{pages}{283--302}.
\newblock \URLprefix \url{https://doi.org/10.1007/BF00874332},
  \DOIprefix\doi{10.1007/BF00874332}.
%Type = Article
\bibitem[{Garagash and Germanovich(2012)}]{garagash_nucleation_2012}
\bibinfo{author}{Garagash, D.I.}, \bibinfo{author}{Germanovich, L.N.},
  \bibinfo{year}{2012}.
\newblock \bibinfo{title}{Nucleation and arrest of dynamic slip on a
  pressurized fault}.
\newblock \bibinfo{journal}{Journal of Geophysical Research: Solid Earth}
  \bibinfo{volume}{117}.
\newblock \URLprefix
  \url{https://agupubs.onlinelibrary.wiley.com/doi/abs/10.1029/2012JB009209},
  \DOIprefix\doi{10.1029/2012JB009209}.
%Type = Article
\bibitem[{Geubelle and Rice(1995)}]{geubelle_spectral_1995}
\bibinfo{author}{Geubelle, P.H.}, \bibinfo{author}{Rice, J.R.},
  \bibinfo{year}{1995}.
\newblock \bibinfo{title}{A spectral method for three-dimensional elastodynamic
  fracture problems}.
\newblock \bibinfo{journal}{Journal of the Mechanics and Physics of Solids}
  \bibinfo{volume}{43}, \bibinfo{pages}{1791--1824}.
\newblock \URLprefix
  \url{http://www.sciencedirect.com/science/article/pii/002250969500043I},
  \DOIprefix\doi{10.1016/0022-5096(95)00043-I}.
%Type = Article
\bibitem[{Geus et~al.(2019)Geus, Popović, Ji, Rosso and Wyart}]{geus_how_2019}
\bibinfo{author}{Geus, T.W.J.d.}, \bibinfo{author}{Popović, M.},
  \bibinfo{author}{Ji, W.}, \bibinfo{author}{Rosso, A.},
  \bibinfo{author}{Wyart, M.}, \bibinfo{year}{2019}.
\newblock \bibinfo{title}{How collective asperity detachments nucleate slip at
  frictional interfaces}.
\newblock \bibinfo{journal}{Proceedings of the National Academy of Sciences}
  \bibinfo{volume}{116}, \bibinfo{pages}{23977--23983}.
\newblock \URLprefix \url{https://www.pnas.org/content/116/48/23977},
  \DOIprefix\doi{10.1073/pnas.1906551116}.
%Type = Article
\bibitem[{Greenwood et~al.(1966)Greenwood, Williamson and
  Bowden}]{greenwood_contact_1966}
\bibinfo{author}{Greenwood, J.A.}, \bibinfo{author}{Williamson, J.B.P.},
  \bibinfo{author}{Bowden, F.P.}, \bibinfo{year}{1966}.
\newblock \bibinfo{title}{Contact of nominally flat surfaces}.
\newblock \bibinfo{journal}{Proceedings of the Royal Society of London. Series
  A. Mathematical and Physical Sciences} \bibinfo{volume}{295},
  \bibinfo{pages}{300--319}.
\newblock \URLprefix
  \url{https://royalsocietypublishing.org/doi/abs/10.1098/rspa.1966.0242},
  \DOIprefix\doi{10.1098/rspa.1966.0242}.
%Type = Book
\bibitem[{Grigoriu(1995)}]{grigoriu_applied_1995}
\bibinfo{author}{Grigoriu, M.}, \bibinfo{year}{1995}.
\newblock \bibinfo{title}{Applied non-{Gaussian} processes: examples, theory,
  simulation, linear random vibration, and {MATLAB} solutions}.
\newblock \bibinfo{publisher}{PTR Prentice Hall}, \bibinfo{address}{Englewood
  Cliffs, NJ}.
\newblock
  \bibinfo{note}{Http://newcatalog.library.cornell.edu/catalog/2681811}.
%Type = Article
\bibitem[{Hinkle et~al.(2020)Hinkle, Nöhring, Leute, Junge and
  Pastewka}]{hinkle_emergence_2020}
\bibinfo{author}{Hinkle, A.R.}, \bibinfo{author}{Nöhring, W.G.},
  \bibinfo{author}{Leute, R.}, \bibinfo{author}{Junge, T.},
  \bibinfo{author}{Pastewka, L.}, \bibinfo{year}{2020}.
\newblock \bibinfo{title}{The emergence of small-scale self-affine surface
  roughness from deformation}.
\newblock \bibinfo{journal}{Science Advances} \bibinfo{volume}{6},
  \bibinfo{pages}{eaax0847}.
\newblock \URLprefix
  \url{https://advances.sciencemag.org/content/6/7/eaax0847},
  \DOIprefix\doi{10.1126/sciadv.aax0847}.
%Type = Article
\bibitem[{Hyun and Robbins(2007)}]{hyun_elastic_2007}
\bibinfo{author}{Hyun, S.}, \bibinfo{author}{Robbins, M.O.},
  \bibinfo{year}{2007}.
\newblock \bibinfo{title}{Elastic contact between rough surfaces: {Effect} of
  roughness at large and small wavelengths}.
\newblock \bibinfo{journal}{Tribology International} \bibinfo{volume}{40},
  \bibinfo{pages}{1413--1422}.
\newblock \URLprefix
  \url{http://www.sciencedirect.com/science/article/pii/S0301679X07000369},
  \DOIprefix\doi{10.1016/j.triboint.2007.02.003}.
%Type = Article
\bibitem[{Kammer and McLaskey(2019)}]{kammer_fracture_2019}
\bibinfo{author}{Kammer, D.S.}, \bibinfo{author}{McLaskey, G.C.},
  \bibinfo{year}{2019}.
\newblock \bibinfo{title}{Fracture energy estimates from large-scale laboratory
  earthquakes}.
\newblock \bibinfo{journal}{Earth and Planetary Science Letters}
  \bibinfo{volume}{511}, \bibinfo{pages}{36--43}.
\newblock \URLprefix
  \url{http://www.sciencedirect.com/science/article/pii/S0012821X19300573},
  \DOIprefix\doi{10.1016/j.epsl.2019.01.031}.
%Type = Article
\bibitem[{Kammer et~al.(2015)Kammer, Radiguet, Ampuero and
  Molinari}]{kammer_linear_2015}
\bibinfo{author}{Kammer, D.S.}, \bibinfo{author}{Radiguet, M.},
  \bibinfo{author}{Ampuero, J.P.}, \bibinfo{author}{Molinari, J.F.},
  \bibinfo{year}{2015}.
\newblock \bibinfo{title}{Linear {Elastic} {Fracture} {Mechanics} {Predicts}
  the {Propagation} {Distance} of {Frictional} {Slip}}.
\newblock \bibinfo{journal}{Tribology Letters} \bibinfo{volume}{57},
  \bibinfo{pages}{23}.
\newblock \URLprefix \url{https://doi.org/10.1007/s11249-014-0451-8},
  \DOIprefix\doi{10.1007/s11249-014-0451-8}.
%Type = Article
\bibitem[{Ke et~al.(2018)Ke, McLaskey and Kammer}]{ke_rupture_2018}
\bibinfo{author}{Ke, C.Y.}, \bibinfo{author}{McLaskey, G.C.},
  \bibinfo{author}{Kammer, D.S.}, \bibinfo{year}{2018}.
\newblock \bibinfo{title}{Rupture {Termination} in {Laboratory}-{Generated}
  {Earthquakes}}.
\newblock \bibinfo{journal}{Geophysical Research Letters} \bibinfo{volume}{45},
  \bibinfo{pages}{12,784--12,792}.
\newblock \URLprefix
  \url{https://agupubs.onlinelibrary.wiley.com/doi/abs/10.1029/2018GL080492},
  \DOIprefix\doi{10.1029/2018GL080492}.
%Type = Techreport
\bibitem[{Ke et~al.(2019)Ke, McLaskey and Kammer}]{ke_analytical_2019}
\bibinfo{author}{Ke, C.Y.}, \bibinfo{author}{McLaskey, G.C.},
  \bibinfo{author}{Kammer, D.S.}, \bibinfo{year}{2019}.
\newblock \bibinfo{title}{Analytical {Crack} {Model} {Inferred} from
  {Contained} {Laboratory}-{Generated} {Earthquakes}}.
\newblock \bibinfo{type}{preprint} \bibinfo{number}{DOI:
  10.31223/osf.io/ur2pf}. EarthArXiv.
\newblock \URLprefix \url{https://osf.io/ur2pf}.
%Type = Article
\bibitem[{Latour et~al.(2013)Latour, Schubnel, Nielsen, Madariaga and
  Vinciguerra}]{latour_characterization_2013}
\bibinfo{author}{Latour, S.}, \bibinfo{author}{Schubnel, A.},
  \bibinfo{author}{Nielsen, S.}, \bibinfo{author}{Madariaga, R.},
  \bibinfo{author}{Vinciguerra, S.}, \bibinfo{year}{2013}.
\newblock \bibinfo{title}{Characterization of nucleation during laboratory
  earthquakes}.
\newblock \bibinfo{journal}{Geophysical Research Letters} \bibinfo{volume}{40},
  \bibinfo{pages}{5064--5069}.
\newblock \URLprefix
  \url{https://agupubs.onlinelibrary.wiley.com/doi/abs/10.1002/grl.50974},
  \DOIprefix\doi{10.1002/grl.50974}.
%Type = Article
\bibitem[{Li and Kim(2008)}]{li_micromechanics_2008}
\bibinfo{author}{Li, Q.}, \bibinfo{author}{Kim, K.S.}, \bibinfo{year}{2008}.
\newblock \bibinfo{title}{Micromechanics of friction: effects of
  nanometre-scale roughness}.
\newblock \bibinfo{journal}{Proceedings of the Royal Society A: Mathematical,
  Physical and Engineering Sciences} \bibinfo{volume}{464},
  \bibinfo{pages}{1319--1343}.
\newblock \URLprefix
  \url{https://royalsocietypublishing.org/doi/10.1098/rspa.2007.0364},
  \DOIprefix\doi{10.1098/rspa.2007.0364}.
%Type = Article
\bibitem[{Li et~al.(2013)Li, Popov, Dimaki, Filippov, Kürschner and
  Popov}]{li_friction_2013}
\bibinfo{author}{Li, Q.}, \bibinfo{author}{Popov, M.}, \bibinfo{author}{Dimaki,
  A.}, \bibinfo{author}{Filippov, A.E.}, \bibinfo{author}{Kürschner, S.},
  \bibinfo{author}{Popov, V.L.}, \bibinfo{year}{2013}.
\newblock \bibinfo{title}{Friction {Between} a {Viscoelastic} {Body} and a
  {Rigid} {Surface} with {Random} {Self}-{Affine} {Roughness}}.
\newblock \bibinfo{journal}{Physical Review Letters} \bibinfo{volume}{111},
  \bibinfo{pages}{034301}.
\newblock \URLprefix
  \url{https://link.aps.org/doi/10.1103/PhysRevLett.111.034301},
  \DOIprefix\doi{10.1103/PhysRevLett.111.034301}.
%Type = Article
\bibitem[{Li et~al.(2018)Li, Pastewka and Szlufarska}]{li_chemical_2018}
\bibinfo{author}{Li, Z.}, \bibinfo{author}{Pastewka, L.},
  \bibinfo{author}{Szlufarska, I.}, \bibinfo{year}{2018}.
\newblock \bibinfo{title}{Chemical aging of large-scale randomly rough
  frictional contacts}.
\newblock \bibinfo{journal}{Physical Review E} \bibinfo{volume}{98},
  \bibinfo{pages}{023001}.
\newblock \URLprefix \url{https://link.aps.org/doi/10.1103/PhysRevE.98.023001},
  \DOIprefix\doi{10.1103/PhysRevE.98.023001}.
%Type = Article
\bibitem[{Lyashenko et~al.(2013)Lyashenko, Pastewka and
  Persson}]{lyashenko_comment_2013}
\bibinfo{author}{Lyashenko, I.A.}, \bibinfo{author}{Pastewka, L.},
  \bibinfo{author}{Persson, B.N.J.}, \bibinfo{year}{2013}.
\newblock \bibinfo{title}{Comment on ``{Friction} {Between} a {Viscoelastic}
  {Body} and a {Rigid} {Surface} with {Random} {Self}-{Affine} {Roughness}''}.
\newblock \bibinfo{journal}{Physical Review Letters} \bibinfo{volume}{111},
  \bibinfo{pages}{189401}.
\newblock \URLprefix
  \url{https://link.aps.org/doi/10.1103/PhysRevLett.111.189401},
  \DOIprefix\doi{10.1103/PhysRevLett.111.189401}.
%Type = Article
\bibitem[{McLaskey(2019)}]{mclaskey_earthquake_2019}
\bibinfo{author}{McLaskey, G.C.}, \bibinfo{year}{2019}.
\newblock \bibinfo{title}{Earthquake {Initiation} {From} {Laboratory}
  {Observations} and {Implications} for {Foreshocks}}.
\newblock \bibinfo{journal}{Journal of Geophysical Research: Solid Earth}
  \bibinfo{volume}{124}, \bibinfo{pages}{12882--12904}.
\newblock \URLprefix
  \url{https://agupubs.onlinelibrary.wiley.com/doi/abs/10.1029/2019JB018363},
  \DOIprefix\doi{10.1029/2019JB018363}.
%Type = Article
\bibitem[{Noda et~al.(2013)Noda, Nakatani and Hori}]{noda_large_2013}
\bibinfo{author}{Noda, H.}, \bibinfo{author}{Nakatani, M.},
  \bibinfo{author}{Hori, T.}, \bibinfo{year}{2013}.
\newblock \bibinfo{title}{Large nucleation before large earthquakes is
  sometimes skipped due to cascade-up—{Implications} from a rate and state
  simulation of faults with hierarchical asperities}.
\newblock \bibinfo{journal}{Journal of Geophysical Research: Solid Earth}
  \bibinfo{volume}{118}, \bibinfo{pages}{2924--2952}.
\newblock \URLprefix
  \url{http://agupubs.onlinelibrary.wiley.com/doi/abs/10.1002/jgrb.50211},
  \DOIprefix\doi{10.1002/jgrb.50211}. \bibinfo{note}{\_eprint:
  https://onlinelibrary.wiley.com/doi/pdf/10.1002/jgrb.50211}.
%Type = Article
\bibitem[{Pei et~al.(2005)Pei, Hyun, Molinari and Robbins}]{pei_finite_2005}
\bibinfo{author}{Pei, L.}, \bibinfo{author}{Hyun, S.},
  \bibinfo{author}{Molinari, J.F.}, \bibinfo{author}{Robbins, M.O.},
  \bibinfo{year}{2005}.
\newblock \bibinfo{title}{Finite element modeling of elasto-plastic contact
  between rough surfaces}.
\newblock \bibinfo{journal}{Journal of the Mechanics and Physics of Solids}
  \bibinfo{volume}{53}, \bibinfo{pages}{2385--2409}.
\newblock \URLprefix
  \url{http://www.sciencedirect.com/science/article/pii/S0022509605001225},
  \DOIprefix\doi{10.1016/j.jmps.2005.06.008}.
%Type = Article
\bibitem[{Persson(2001)}]{persson_theory_2001}
\bibinfo{author}{Persson, B.N.J.}, \bibinfo{year}{2001}.
\newblock \bibinfo{title}{Theory of rubber friction and contact mechanics}.
\newblock \bibinfo{journal}{The Journal of Chemical Physics}
  \bibinfo{volume}{115}, \bibinfo{pages}{3840--3861}.
\newblock \URLprefix \url{https://aip.scitation.org/doi/10.1063/1.1388626},
  \DOIprefix\doi{10.1063/1.1388626}.
%Type = Article
\bibitem[{Pilvelait et~al.(2020)Pilvelait, Dillavou and
  Rubinstein}]{pilvelait_influences_2020}
\bibinfo{author}{Pilvelait, T.}, \bibinfo{author}{Dillavou, S.},
  \bibinfo{author}{Rubinstein, S.M.}, \bibinfo{year}{2020}.
\newblock \bibinfo{title}{Influences of microcontact shape on the state of a
  frictional interface}.
\newblock \bibinfo{journal}{Physical Review Research} \bibinfo{volume}{2},
  \bibinfo{pages}{012056}.
\newblock \URLprefix
  \url{https://link.aps.org/doi/10.1103/PhysRevResearch.2.012056},
  \DOIprefix\doi{10.1103/PhysRevResearch.2.012056}.
%Type = Article
\bibitem[{Popova and Popov(2015)}]{popova_research_2015}
\bibinfo{author}{Popova, E.}, \bibinfo{author}{Popov, V.L.},
  \bibinfo{year}{2015}.
\newblock \bibinfo{title}{The research works of {Coulomb} and {Amontons} and
  generalized laws of friction}.
\newblock \bibinfo{journal}{Friction} \bibinfo{volume}{3},
  \bibinfo{pages}{183--190}.
\newblock \URLprefix \url{https://doi.org/10.1007/s40544-015-0074-6},
  \DOIprefix\doi{10.1007/s40544-015-0074-6}.
%Type = Article
\bibitem[{Rabinowicz(1992)}]{rabinowicz_friction_1992}
\bibinfo{author}{Rabinowicz, E.}, \bibinfo{year}{1992}.
\newblock \bibinfo{title}{Friction coefficients of noble metals over a range of
  loads}.
\newblock \bibinfo{journal}{Wear} \bibinfo{volume}{159},
  \bibinfo{pages}{89--94}.
\newblock \URLprefix
  \url{http://www.sciencedirect.com/science/article/pii/004316489290289K},
  \DOIprefix\doi{10.1016/0043-1648(92)90289-K}.
%Type = Book
\bibitem[{Rabinowicz(1995)}]{rabinowicz_friction_1995}
\bibinfo{author}{Rabinowicz, E.}, \bibinfo{year}{1995}.
\newblock \bibinfo{title}{Friction and wear of materials}.
\newblock \bibinfo{edition}{2nd ed} ed., \bibinfo{publisher}{Wiley},
  \bibinfo{address}{New York}.
%Type = Article
\bibitem[{Ray and Viesca(2017)}]{ray_earthquake_2017}
\bibinfo{author}{Ray, S.}, \bibinfo{author}{Viesca, R.C.},
  \bibinfo{year}{2017}.
\newblock \bibinfo{title}{Earthquake {Nucleation} on {Faults} {With}
  {Heterogeneous} {Frictional} {Properties}, {Normal} {Stress}}.
\newblock \bibinfo{journal}{Journal of Geophysical Research: Solid Earth}
  \bibinfo{volume}{122}, \bibinfo{pages}{8214--8240}.
\newblock \URLprefix
  \url{https://agupubs.onlinelibrary.wiley.com/doi/abs/10.1002/2017JB014521},
  \DOIprefix\doi{10.1002/2017JB014521}.
%Type = Article
\bibitem[{Ray and Viesca(2019)}]{ray_homogenization_2019}
\bibinfo{author}{Ray, S.}, \bibinfo{author}{Viesca, R.C.},
  \bibinfo{year}{2019}.
\newblock \bibinfo{title}{Homogenization of fault frictional properties}.
\newblock \bibinfo{journal}{Geophysical Journal International}
  \bibinfo{volume}{219}, \bibinfo{pages}{1203--1211}.
\newblock \URLprefix
  \url{https://academic.oup.com/gji/article/219/2/1203/5533328},
  \DOIprefix\doi{10.1093/gji/ggz327}.
%Type = Article
\bibitem[{Rice and Ruina(1983)}]{rice_stability_1983}
\bibinfo{author}{Rice, J.R.}, \bibinfo{author}{Ruina, A.L.},
  \bibinfo{year}{1983}.
\newblock \bibinfo{title}{Stability of {Steady} {Frictional} {Slipping}}.
\newblock \bibinfo{journal}{Journal of Applied Mechanics} \bibinfo{volume}{50},
  \bibinfo{pages}{343--349}.
\newblock \URLprefix
  \url{https://asmedigitalcollection.asme.org/appliedmechanics/article/50/2/343/389397/Stability-of-Steady-Frictional-Slipping},
  \DOIprefix\doi{10.1115/1.3167042}.
%Type = Article
\bibitem[{Rubin and Ampuero(2005)}]{rubin_earthquake_2005}
\bibinfo{author}{Rubin, A.M.}, \bibinfo{author}{Ampuero, J.P.},
  \bibinfo{year}{2005}.
\newblock \bibinfo{title}{Earthquake nucleation on (aging) rate and state
  faults}.
\newblock \bibinfo{journal}{Journal of Geophysical Research: Solid Earth}
  \bibinfo{volume}{110}.
\newblock \URLprefix
  \url{https://agupubs.onlinelibrary.wiley.com/doi/abs/10.1029/2005JB003686},
  \DOIprefix\doi{10.1029/2005JB003686}.
%Type = Article
\bibitem[{Rubino et~al.(2017)Rubino, Rosakis and
  Lapusta}]{rubino_understanding_2017}
\bibinfo{author}{Rubino, V.}, \bibinfo{author}{Rosakis, A.J.},
  \bibinfo{author}{Lapusta, N.}, \bibinfo{year}{2017}.
\newblock \bibinfo{title}{Understanding dynamic friction through spontaneously
  evolving laboratory earthquakes}.
\newblock \bibinfo{journal}{Nature Communications} \bibinfo{volume}{8},
  \bibinfo{pages}{1--13}.
\newblock \URLprefix \url{https://www.nature.com/articles/ncomms15991/},
  \DOIprefix\doi{10.1038/ncomms15991}.
%Type = Article
\bibitem[{Rubinstein et~al.(2007)Rubinstein, Cohen and
  Fineberg}]{rubinstein_dynamics_2007}
\bibinfo{author}{Rubinstein, S.M.}, \bibinfo{author}{Cohen, G.},
  \bibinfo{author}{Fineberg, J.}, \bibinfo{year}{2007}.
\newblock \bibinfo{title}{Dynamics of {Precursors} to {Frictional} {Sliding}}.
\newblock \bibinfo{journal}{Physical Review Letters} \bibinfo{volume}{98},
  \bibinfo{pages}{226103}.
\newblock \URLprefix
  \url{https://link.aps.org/doi/10.1103/PhysRevLett.98.226103},
  \DOIprefix\doi{10.1103/PhysRevLett.98.226103}.
%Type = Article
\bibitem[{Ruina(1983)}]{ruina_slip_1983}
\bibinfo{author}{Ruina, A.}, \bibinfo{year}{1983}.
\newblock \bibinfo{title}{Slip instability and state variable friction laws}.
\newblock \bibinfo{journal}{Journal of Geophysical Research: Solid Earth}
  \bibinfo{volume}{88}, \bibinfo{pages}{10359--10370}.
\newblock \URLprefix
  \url{https://agupubs.onlinelibrary.wiley.com/doi/abs/10.1029/JB088iB12p10359},
  \DOIprefix\doi{10.1029/JB088iB12p10359}.
%Type = Article
\bibitem[{Sahli et~al.(2018)Sahli, Pallares, Ducottet, Ali, Akhrass, Guibert
  and Scheibert}]{sahli_evolution_2018}
\bibinfo{author}{Sahli, R.}, \bibinfo{author}{Pallares, G.},
  \bibinfo{author}{Ducottet, C.}, \bibinfo{author}{Ali, I.E.B.},
  \bibinfo{author}{Akhrass, S.A.}, \bibinfo{author}{Guibert, M.},
  \bibinfo{author}{Scheibert, J.}, \bibinfo{year}{2018}.
\newblock \bibinfo{title}{Evolution of real contact area under shear and the
  value of static friction of soft materials}.
\newblock \bibinfo{journal}{Proceedings of the National Academy of Sciences}
  \bibinfo{volume}{115}, \bibinfo{pages}{471--476}.
\newblock \URLprefix \url{https://www.pnas.org/content/115/3/471},
  \DOIprefix\doi{10.1073/pnas.1706434115}.
%Type = Article
\bibitem[{Savio et~al.(2016)Savio, Pastewka and Gumbsch}]{savio_boundary_2016}
\bibinfo{author}{Savio, D.}, \bibinfo{author}{Pastewka, L.},
  \bibinfo{author}{Gumbsch, P.}, \bibinfo{year}{2016}.
\newblock \bibinfo{title}{Boundary lubrication of heterogeneous surfaces and
  the onset of cavitation in frictional contacts}.
\newblock \bibinfo{journal}{Science Advances} \bibinfo{volume}{2},
  \bibinfo{pages}{e1501585}.
\newblock \URLprefix
  \url{https://advances.sciencemag.org/content/2/3/e1501585},
  \DOIprefix\doi{10.1126/sciadv.1501585}.
%Type = Article
\bibitem[{Sokoloff(2001)}]{sokoloff_static_2001}
\bibinfo{author}{Sokoloff, J.B.}, \bibinfo{year}{2001}.
\newblock \bibinfo{title}{Static {Friction} between {Elastic} {Solids} due to
  {Random} {Asperities}}.
\newblock \bibinfo{journal}{Physical Review Letters} \bibinfo{volume}{86},
  \bibinfo{pages}{3312--3315}.
\newblock \URLprefix
  \url{https://link.aps.org/doi/10.1103/PhysRevLett.86.3312},
  \DOIprefix\doi{10.1103/PhysRevLett.86.3312}.
%Type = Book
\bibitem[{Spencer and Tysoe(2015)}]{spencer_cutting_2015}
\bibinfo{author}{Spencer, N.D.}, \bibinfo{author}{Tysoe, W.T.},
  \bibinfo{year}{2015}.
\newblock \bibinfo{title}{The cutting edge of tribology: a decade of progress
  in friction, lubrication, and wear}.
\newblock \bibinfo{publisher}{World Scientific},
  \bibinfo{address}{[Hackensack], New Jersey}.
\newblock \bibinfo{note}{OCLC: ocn907104207}.
%Type = Article
\bibitem[{Svetlizky et~al.(2020)Svetlizky, Albertini, Cohen, Kammer and
  Fineberg}]{svetlizky_dynamic_2020}
\bibinfo{author}{Svetlizky, I.}, \bibinfo{author}{Albertini, G.},
  \bibinfo{author}{Cohen, G.}, \bibinfo{author}{Kammer, D.S.},
  \bibinfo{author}{Fineberg, J.}, \bibinfo{year}{2020}.
\newblock \bibinfo{title}{Dynamic fields at the tip of sub-{Rayleigh} and
  supershear frictional rupture fronts}.
\newblock \bibinfo{journal}{Journal of the Mechanics and Physics of Solids}
  \bibinfo{volume}{137}, \bibinfo{pages}{103826}.
\newblock \URLprefix
  \url{http://www.sciencedirect.com/science/article/pii/S0022509619308282},
  \DOIprefix\doi{10.1016/j.jmps.2019.103826}.
%Type = Article
\bibitem[{Svetlizky and Fineberg(2014)}]{svetlizky_classical_2014}
\bibinfo{author}{Svetlizky, I.}, \bibinfo{author}{Fineberg, J.},
  \bibinfo{year}{2014}.
\newblock \bibinfo{title}{Classical shear cracks drive the onset of dry
  frictional motion}.
\newblock \bibinfo{journal}{Nature} \bibinfo{volume}{509},
  \bibinfo{pages}{205--208}.
\newblock \URLprefix \url{https://www.nature.com/articles/nature13202},
  \DOIprefix\doi{10.1038/nature13202}.
%Type = Book
\bibitem[{Thomas(1999)}]{thomas_rough_1999}
\bibinfo{author}{Thomas, T.R.}, \bibinfo{year}{1999}.
\newblock \bibinfo{title}{Rough surfaces}.
\newblock \bibinfo{edition}{2. ed} ed., \bibinfo{publisher}{Imperial College
  Press}, \bibinfo{address}{London}.
\newblock \bibinfo{note}{OCLC: 632713233}.
%Type = Article
\bibitem[{Uenishi and Rice(2003)}]{uenishi_universal_2003}
\bibinfo{author}{Uenishi, K.}, \bibinfo{author}{Rice, J.R.},
  \bibinfo{year}{2003}.
\newblock \bibinfo{title}{Universal nucleation length for slip-weakening
  rupture instability under nonuniform fault loading}.
\newblock \bibinfo{journal}{Journal of Geophysical Research: Solid Earth}
  \bibinfo{volume}{108}.
\newblock \URLprefix
  \url{https://agupubs.onlinelibrary.wiley.com/doi/abs/10.1029/2001JB001681},
  \DOIprefix\doi{10.1029/2001JB001681}.
%Type = Article
\bibitem[{Yastrebov et~al.(2015)Yastrebov, Anciaux and
  Molinari}]{yastrebov_infinitesimal_2015}
\bibinfo{author}{Yastrebov, V.A.}, \bibinfo{author}{Anciaux, G.},
  \bibinfo{author}{Molinari, J.F.}, \bibinfo{year}{2015}.
\newblock \bibinfo{title}{From infinitesimal to full contact between rough
  surfaces: {Evolution} of the contact area}.
\newblock \bibinfo{journal}{International Journal of Solids and Structures}
  \bibinfo{volume}{52}, \bibinfo{pages}{83--102}.
\newblock \URLprefix
  \url{http://www.sciencedirect.com/science/article/pii/S0020768314003667},
  \DOIprefix\doi{10.1016/j.ijsolstr.2014.09.019}.

\end{thebibliography}

\appendix
\section{Nucleation Criterion}
\label{sec:appendixA}

The nucleation criterion used in this work is based on the theory developed by \citet{uenishi_universal_2003}. It is not our intention of re-deriving the theoretical framework. Nevertheless, in this section, we provide a clear problem statement such that our work can easily be related to the work by \citet{uenishi_universal_2003}.
The peak strength along the interface is given by
\begin{equation}
    \taup(x) = \taupmin + q(x) ~,
    \label{eq:apxAtaup}
\end{equation}
where $\taupmin$ is the minimum value of $\taup(x)$. The functional form $q(x)$ satisfies $q(x_m) = 0$ and $q(x) > 0$ for $x \neq x_m$. If local slip occurs at any point along the interface, the local strength decreases because of the slip-weakening friction law, as defined by Eq.~\ref{eq:linslipweaklaw}. Therefore, any point that is in the weakening process, \textit{i.e.}, $\dc > \slip(x,t) > 0$, presents a local shear stress that is given by
\begin{equation}
    \tau(x) = \taup(x) - \weakrate \slip(x,t) = \taupmin + q(x) - \weakrate \slip(x,t) ~,
    \label{eq:apxAtau}
\end{equation}
where Eq.~\ref{eq:apxAtaup} was used and the weakening rate satisfies $\weakrate>0$. 

The applied shear stress, which starts at the level of the minimum strength, is defined by
\begin{equation}
    \taui(t) = \taupmin + Rt ~,
    \label{eq:apxAtaui}
\end{equation}
where $R>0$ is the shear-stress loading rate.

Following \citet{uenishi_universal_2003}, we can consider the quasi-static elastic equilibrium~\citep{bilby_disclocations_1968} that relates the stress change along the interface with the local slip through
\begin{equation}
    \tau(x,t) = \taui(x,t) - \frac{\equivshearmodulus}{2\pi} \int_{a_-(t)}^{a_+(t)} \frac{\partial \slip(\xi,t)/\partial \xi}{x-\xi} \mathrm{d}\xi ~,
    \label{eq:apxAeshelby}
\end{equation}
where $\equivshearmodulus = \shearmodulus / (1-\poisson)$ and $a_-(t) < x < a_+(t)$ are the boundaries of the slowly expanding slipping area. By substituting Eq.~\ref{eq:apxAtau} and Eq.~\ref{eq:apxAtaui} into Eq.~\ref{eq:apxAeshelby}, we find
\begin{equation}
    - \weakrate \slip(x,t) = Rt - q(x) - \frac{\equivshearmodulus}{2\pi} \int_{a_-(t)}^{a_+(t)} \frac{\partial \slip(\xi,t)/\partial \xi}{x-\xi} \mathrm{d}\xi ~,
    \label{eq:apxAprobstat}
\end{equation}
for $\slip(x,t)>0$ and $a_-(t) < x < a_+(t)$. This corresponds exactly to \citep[Eq.4]{uenishi_universal_2003}.

Starting from this equation, \citet{uenishi_universal_2003} show that quasi-static solutions cease to exist for slipping areas larger than a critical length $\hn$, which is given by
\begin{equation}
    \hn \approx 1.158 \frac{\equivshearmodulus}{\weakrate} ~.
    \label{eq:apxAcritleng}
\end{equation}
Interestingly, the critical length only depends on the shear modulus $\equivshearmodulus$ and the slip-weakening rate $\weakrate$, and is independent of the loading rate $R$ and the shape of the peak strength $q(x)$.

\citet{uenishi_universal_2003} further show that a slipping area exceeding $\hn$ is reached at time $t_c$ when the critical stress level is given by \citep[Eq.14]{uenishi_universal_2003}
\begin{equation}
    Rt_c \approx 0.751 \int_{-1}^{+1} q[a(t_c)s + b(t_c)] v_0(s) \mathrm{d}s ~,
    \label{eq:apxAcrittau}
\end{equation}
where $a(t) = [a_+(t) - a_-(t)]/2$ and $b(t) = [a_+(t) + a_-(t)]/2$ are the half-length and center location of the slipping area, respectively, and $s=[x-b(t)]/a(t)$ and $v_0(s) \approx (0.925 - 0.308s^2) \sqrt{1-s^2}$. It becomes obvious that the stress level at which the slipping area reaches the critical length depends on the shape of $q(x)$.

%---------------------

\section{Simplified Statistical Analysis of the Nucleation Strength}\label{sec:appendixStats}

In order to give some intuition of the effects of correlation length $\corlen$ on the nucleation strength $\taun$ (Eq.~\ref{eq:taun}), we provide a statistical argument, which is based on the property of stationarity of $\taup$. Note that $v_0(.)$ has the following property
\begin{equation}
    0.751\int_{-1}^{+1} v_0(s)\mathrm{d}s=1
\end{equation}

\begin{figure}
\begin{center}
  \includegraphics[width=3.5in]{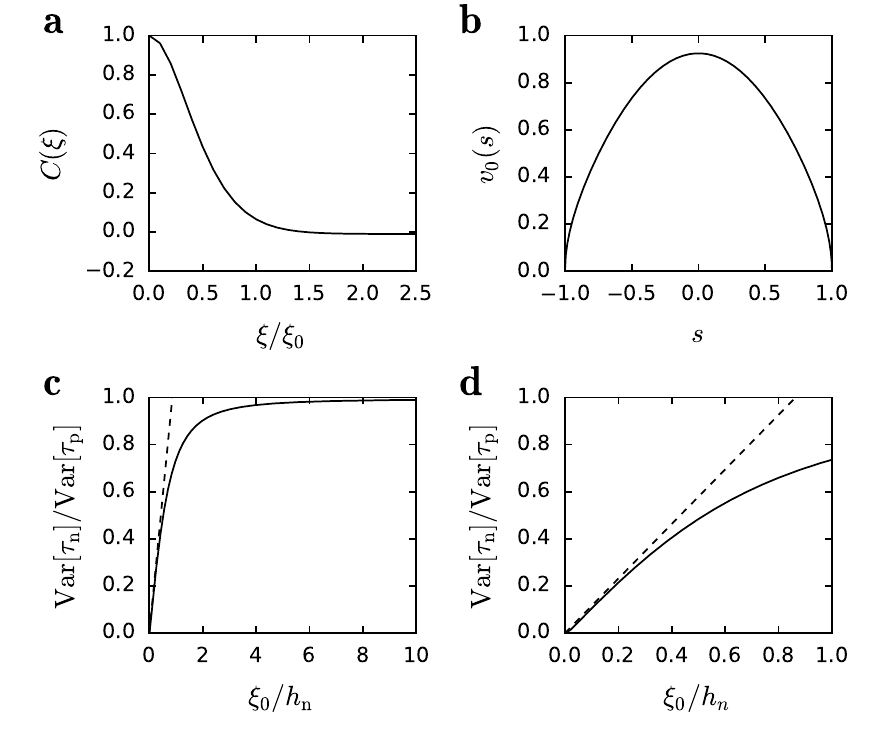}
\caption{Numerical evaluation of Eq.\ref{eq:AnnexBVar}. (a)~Correlation function. (b)~First eigenfunction of the elastic problem. (c)~Normalized variance of the nucleation strength $\taun$. (d)~Zoom over $\corlen<\hn$. Dashed line in (c,d) represents }
  \label{fig:variance}  
\end{center}
\end{figure}

We aim to evaluate the expectation and variance of $\taun$ as function of $\corlen$. The expectation is an integral with respect to a probability measure rather than a Lebesgue measure. 
Since $\taup$ and $\taun$ are stationary, we can apply the Fubini's theorem, which states that the order of integration can be changed, and express the expectation $\E[\taun]$ as function of the expectation of the local strength $\E[\taup]$.

\begin{equation}\label{eq:AnnexBExp}
\begin{split}
\E[\taun]&= \E\left[0.751\int_{-1}^{+1} \taup\left(s\,\hn/2+x\right) v_0(s) \mathrm{d}s\right]=0.751\int_{-1}^{+1} \E\left[\taup\left(s\,\hn/2+x\right)\right]v_0(s)\mathrm{d}s\\
&=\E[\taup] 0.751\int_{-1}^{+1}v_0(s)\mathrm{d}s=\E[\taup]
\end{split}
\end{equation}
Similarly, we can express its variance $\Var[\taun]$ as function of the variance of the local strength $\Var[\taup]$ by applying Fubini's Theorem and the definition of the correlation function $C(\xi)=\E[(\taup(x)-\E[\taup])(\taup(x+\xi)-\E[\taup])/\Var[\taup]$
\begin{equation}\label{eq:AnnexBVar}
\begin{split}
    \Var[\taun] &= \E[(\taun(x) - \E[\taun])^2]=\E\left[\left(0.751 \int_{-1}^{+1} \taup\left(s\,\hn/2+x\right)v_0(s) \mathrm{d}s - \E[\taup]\right)^2\right]\\
    &=\E\left[\left(0.751 \int_{-1}^{+1} \left(\taup\left(s\,\hn/2+x\right)-\E[\taup]\right)v_0(s) \mathrm{d}s \right)^2\right]\\
    &=0.751^2 \iint_{[-1,1]^2} \E\left[\left(\taup\left(s\,\hn/2 +x\right) -\E[\taup]\right)\left(\taup\left(t\,\hn/2+x\right)-\E[\taup]\right)\right] v_0(s)v_0(t) \mathrm{d}s \, \mathrm{d}t    \\
    &= \Var[\taup] 0.751^2 \iint_{[-1,1]^2} C((s-t)\hn/2)v_0(s)v_0(t) \mathrm{d}s\, \mathrm{d}t
\end{split}
\end{equation}
For the limiting cases the expression for the variance can be expressed analytically.
For perfectly correlated $\taup$, $\corlen=\infty$, $C(.)=1$
\begin{equation}\label{eq:ABVarinf}
    \lim_{\corlen/\hn\rightarrow\infty}\Var[\taun]= \Var[\taup]
\end{equation}
Both $C(.)$ and $v_0(.)$ are known. Therefore, the integral of Eq.~\ref{eq:AnnexBVar} can be solved numerically (see Fig.\ref{fig:variance}).
For $\corlen\ll \hn$ the correlation function $C(.) \approx$  Dirac-$\delta$ and the double integral collapses to a single integral.
\begin{equation}\label{eq:ABVarll}
\begin{split}
    {\corlen\ll\hn}\Rightarrow \Var[\taun] \propto {\Var[\taup]} \int_{-1}^{+1} \frac{\corlen}{\hn} v_0^2(s) \mathrm{d} s
    \propto {\Var[\taup]}\frac{\corlen}{\hn}
\end{split}
\end{equation}
Note the linear scaling for $\corlen\ll\hn$ in Fig.~\ref{fig:variance}d.
\end{document}